\documentclass[doublespace,english,]{revtex4-1}
\usepackage[T1]{fontenc}
\pdfoutput=1
\usepackage{array}
\usepackage{booktabs}
\usepackage{multirow}

\usepackage{amsmath, amsthm, amssymb, amsfonts, enumerate}
\usepackage{setspace}
\usepackage{natbib, bbm}
\usepackage{graphicx}
\usepackage{color}

\providecommand{\tabularnewline}{\\}

\usepackage{babel}

\usepackage{subfig}

\begin{document}

\title{A Flux-Balanced Fluid Model for Collisional Plasma Edge Turbulence: Numerical Simulations with Different Aspect Ratios}

\author{Di Qi, Andrew J. Majda}

\affiliation{Department of Mathematics, Courant Institute
of Mathematical Sciences, New York University, New York, NY 10012}

\affiliation{Center for Atmosphere and Ocean Science, Courant Institute of Mathematical Sciences, New York University, New York, NY 10012}

\author{Antoine J. Cerfon}
\affiliation{Department of Mathematics, Courant Institute
of Mathematical Sciences, New York University, New York, NY 10012}

\begin{abstract}
We investigate the drift wave -- zonal flow dynamics in a shearless slab geometry with the new flux-balanced Hasegawa-Wakatani model. As in previous Hasegawa-Wakatani models, we observe a sharp transition from a turbulence dominated regime to a zonal jet dominated regime as we decrease the plasma resistivity. However, unlike previous models, zonal structures are always present in the flux-balanced model, even for high resistivity, and strongly reduce the level of particle and vorticity flux. The more robust zonal jets also have a higher variability than in previous models, which is further enhanced when the computational domain is chosen to be elongated in the radial direction. In these cases, we observe complex multi-scale dynamics, with multiple jets interacting with one another, and intermittent bursts. We present a detailed statistical analysis which highlights how the changes in the aspect ratio of the computational domain affect the third-order statistical moments, and thus modify the turbulent dynamics.
\end{abstract}

\maketitle

\section{Introduction}

In magnetic confinement devices, the interplay between zonal flows and drift-wave turbulence driven by temperature and density gradients is thought to play a critical role in the observed level of heat and particle transport perpendicular to the magnetic surfaces \citep{hortonreview1999,diamond2005zonal,fujizawa2009,xanthopoulos2011,manz2013}.  Reduced fluid models based on simplifying assumptions for the magnetic geometry and the plasma regime of interest have contributed to improve our understanding of the drift wave -- zonal flow dynamics. Two families of reduced models stand out in that context: the Hasegawa-Mima (HM) models \citep{hasegawa1978pseudo,dewar2007zonal} and the Hasegawa-Wakatani (HW) models \citep{hasegawa1983plasma,numata2007bifurcation}. The HM models are the simplest known models containing the drift wave turbulence -- zonal flow feedback loop mechanism. However, they do not contain any instability, so that turbulent forcing must be added externally in order to study the turbulence -- zonal flow dynamics. The HW models are more complicated, in the sense that they describe the coupled evolution of two fields, namely the electrostatic potential perturbation and the density perturbation, instead of just the electrostatic potential as in the HM models. On the other hand, the HW models are more physically satisfying because they inherently contain a drift instability driving the turbulence, and this turbulence is found to self-organise into zonal structures in certain physical regimes \citep{numata2007bifurcation,pushkarev2013}. 

Even so, the existing HW models are not entirely satisfying. In the original Hasegawa-Wakatani model \citep{hasegawa1983plasma}, zonal components have a net contribution to the parallel current density, which is unphysical in toroidal geometries \citep{numata2007bifurcation}, and inhibits the formation of zonal jets \citep{numata2007bifurcation,pushkarev2013}. The modified HW model (mHW) proposed by Numata \textit{et al.} \citep{numata2007bifurcation} properly removes the contributions of the zonal components to the parallel density, and can capture the transition from a turbulence dominated regime to self-organized regime with strong zonal structures. However, a weakness of the mHW model is that it does not correctly converge to the corresponding HM model, often called the modified HM model (mHM) \citep{dewar2007zonal}, when one takes the limit of adiabatic electrons. A very small plasma resistivity is sufficient to lead to a significant net radial flux of electrons, contrary to what one would physically expect in a toroidal magnetic geometry \citep{dorland1993gyrofluid,hammett1993,dewar2007zonal}. We have recently proposed a new \emph{flux-balanced Hasegawa-Wakatani} (bHW) model with an improved treatment of the electron dynamics parallel to the magnetic field lines which addresses this issue \citep{majda2018flux}. This new reduced fluid model has several desirable properties: it is Galilean invariant under boosts in the poloidal direction; it converges exactly to the modified HM model \citep{dewar2007zonal} in the appropriate limit in which the electrons are adiabatic (which ties the density perturbation to the electrostatic potential perturbation and reduces the model to a one-field theory), it has the same drift instability as the previous HW models, but in comparison to these models, stronger interactions between the turbulent flow and zonal jets are induced in the bHW model, so the zonal jets and fluctuations are greatly enhanced \citep{majda2018flux}. 

\begin{figure}
\begin{centering}
\includegraphics[height=5.cm]{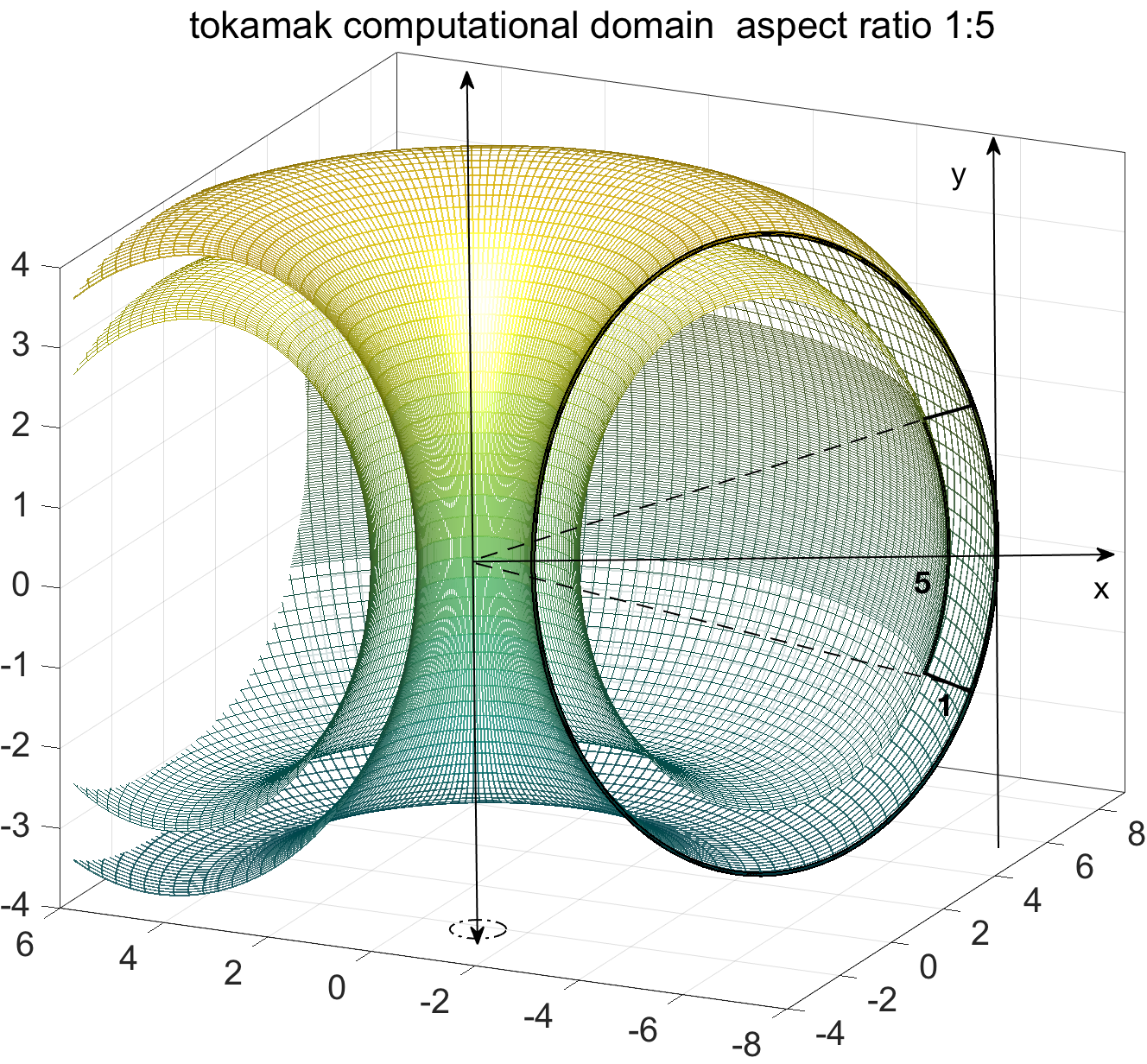}\qquad\includegraphics[height=5.cm]{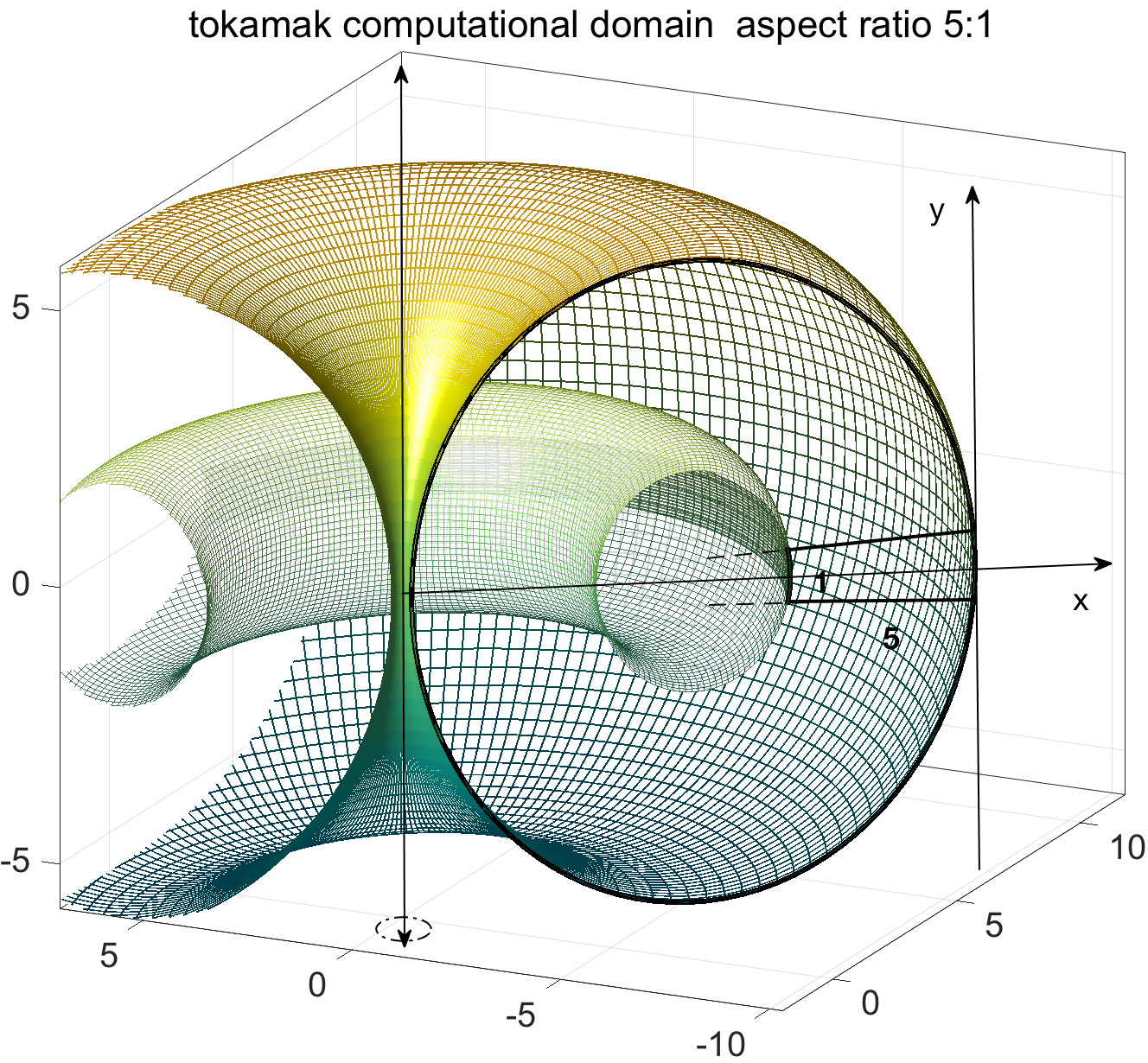}
\par\end{centering}

\centering{}\caption{Illustration of the coordinate system in the plasma computational
domain with aspect ratios 1:5 and 5:1.\label{fig:Illustration}}
\end{figure}

The purpose of the present paper is to look at the mechanisms involved in the drift wave -- zonal flow interaction under the new light of the bHW model. To do so, we will rely on direct numerical simulations, and focus on two specific aspects. First, we will continue the direct comparisons with the well-known mHW model \citep{numata2007bifurcation} we initiated in a previous article \citep{majda2018flux}, in order to further identify the differences due the improved treatment of the electron dynamics in the bHW model. Second, we will investigate the role of the dimensions of the computational domain in the observed dynamics. Direct numerical simulations of the HW equations are usually done within the framework of the flux-tube approximation \citep{beer1995}, in which the boundary conditions for the coordinate corresponding to the poloidal direction, the $y$ coordinate in this article, and the coordinate corresponding to the radial direction, the $x$ coordinate in our work, are both periodic. While the periodicity in the poloidal direction can be intuitively understood, the radial periodicity is less straightforward. The justification usually given for these simple boundary conditions is that if the length of the computational domain in the radial direction is longer than the radial turbulence correlation length, the turbulence is expected to be insensitive to the boundary conditions \citep{beer1995}. The aspect ratio for the computational domain of HW simulations with periodic boundary conditions is therefore effectively determined by the expected ratio of turbulence correlation lengths between the poloidal direction and the radial direction. In this work, we will consider three representative cases: a domain with 1:1 aspect ratio, which corresponds to the standard setup for turbulence simulations \citep{beer1995,parker1999,numata2007bifurcation}, a domain with 1:5 aspect ratio, i.e. elongated in the poloidal direction, and a domain with 5:1 aspect ratio, i.e. elongated in the radial direction. For illustration purposes, the two elongated domains are shown in a tokamak geometry in Figure \ref{fig:Illustration}, although in this article we will consider a simpler shearless slab geometry, as is commonly done for these studies \citep{numata2007bifurcation,majda2018flux}. We will show that the aspect ratio of the computational domain has a profound effect on the observed turbulent dynamics and zonal flow structures. To do so, we will rely on direct numerial simulations to visualise the vorticity and flow fields and to compute the key statistical properties of the turbulence.

The paper is organized in the following way. We first present a brief review of the new flux-balanced Hasegawa-Wakatani (bHW) model in Section \ref{sec:review}. Following this review, we highlight some fundamental features of the drift wave -- zonal flow dynamics observed in the bHW model in Section \ref{sec:Flow-Features}, comparing results from the bHW model with analogous results from the mHW model and focusing on the transition from the drift-wave turbulence dominated regime to the zonal jet dominated regime. This transition is associated with a decrease of the parallel plasma resistivity. In Section \ref{sec:Aspect-ratios} we investigate the effects of the aspect ratio of the computational domain, comparing direct numerical simulations obtained with aspect ratios 5:1 and 1:5 as displayed in Figure \ref{fig:Illustration} together with simulations on a square domain. In Section \ref{sec:Higher-Order-Feedbacks}, we analyse the exchange of energy between the zonal jets and drift waves from a statistical point of view, by considering the dynamics for the first and second order moments and characterizing the role of the higher-order feedbacks. We summarise our work in Section \ref{sec:Summary} and suggest directions for future work.

\section{The Flux-Balanced Hasegawa-Wakatani Model\label{sec:review}}

In this section, we present a brief review of the mHW model \citep{numata2007bifurcation} and of the bHW model introduced in \citep{majda2018flux}, which are the two models with which we will study the drift wave -- zonal flow dynamics throughout this article. 

\subsection{The modified Hasegawa-Wakatani model}

The \emph{ }Hasegawa-Wakatani models\emph{ } all describe the
drift wave -- zonal flow interactions with a coupled system of partial differential equations for two scalar fields in a shearless slab geometry, where the constant and uniform magnetic field is along the $z$-direction, and the physical quantities only depend on the $x$ and $y$ coordinates, the $x$-axis corresponding to the radial
direction and $y$ representing the poloidal direction. In this slab geometry, zonal flows are along the $y$-direction. The original Hasegawa-Wakatani
model (oHW) was introduced in \citep{hasegawa1983plasma} and a modification to the model was proposed in \citep{numata2007bifurcation} when it was recognised that the zonal components should not contribute to the resistive coupling term in the oHW model. This modification is now known as the mHW model, and has been found to lead to the generation of stronger and more ubiquitous zonal flows than in the oHW model. The mHW model is usually formulated in terms of the ion vorticity $\zeta=\Delta\varphi$ and the density perturbation $n=n_{1}/n_{0}$, where $n_{0}$ is the equilibrium density and $N=n_{0}+n_{1}$ is the total density. The mHW model is then given by the following coupled system of partial differential equations
\addtocounter{equation}{0}\begin{subequations}\label{plasma}
\begin{eqnarray}
\frac{\partial\zeta}{\partial t}+J\left(\varphi,\zeta\right) & = & \alpha\left(\tilde{\varphi}-\tilde{n}\right)+D\Delta\zeta,\label{eq:plasma1}\\
\frac{\partial n}{\partial t}+J\left(\varphi,n\right)+\kappa\frac{\partial\varphi}{\partial y} & = & \alpha\left(\tilde{\varphi}-\tilde{n}\right)+D\Delta n.\label{eq:plasma2}
\end{eqnarray}
\end{subequations}
where $J\left(\varphi,f\right)=\partial_{x}\varphi\partial_{y}f-\partial_{y}\varphi\partial_{x}f$ for $f=\zeta, n$, $t$ is the time normalised to the ion cyclotron frequency $\omega_{ci}=eB_{0}/m$, $x$ and $y$ are normalised in terms of the hybrid ion thermal Larmor radius $\rho_{s}=\omega_{\mathrm{ci}}^{-1}\left(T_{e}/m\right)^{1/2}=\sqrt{mT_{e}}/eB_{0}$, and $\varphi=e\phi/T_{e}$ is the normalised electrostatic potential, where $e$ is the charge of the electron and $\phi$ the electrostatic potential. We treat the density profile within the framework of the standard ``local approximation", in which
$n_{0}'(x)/n_{0}(x)=$constant \citep{numata2007bifurcation}, and have introduced $\kappa=-\mathrm{d}\ln n_{0}/\mathrm{d}x$.

$\alpha$ is often called the ``adiabaticity parameter" \citep{numata2007bifurcation} and is proportional to the inverse of the parallel resistivity. The limit of adiabatic electrons is obtained for $\alpha\rightarrow\infty$, while the opposite limit, $\alpha\rightarrow 0$ is the hydrodynamic limit.  $D$ acts on the vorticity and density equations in the Laplace operator as a homogeneous dissipation. Finally, we have decomposed the physical quantities $\varphi$ and $n$ as the sums $\varphi=\overline{\varphi}+\tilde{\varphi}$ and $n=\overline{n}+\tilde{n}$ of their zonal mean states $\overline{\varphi},\overline{n}$ and their fluctuations $\tilde{\varphi},\tilde{n}$, where, for any quantity $f$,
\[
\overline{f}\left(x\right)=\frac{1}{L_{y}}\int f\left(x,y\right)dy,\quad\tilde{f}=f-\overline{f}.
\]
As mentioned above, the difference between the oHW and mHW models is found in the resistive coupling term, which takes the form $\alpha\left(\varphi-n\right)$ in the oHW model, and $\alpha\left(\tilde{\varphi}-\tilde{n}\right)$ in the mHW model given in (\ref{plasma}). 

The mHW model has been a popular model for the study of the drift wave -- zonal flow dynamics because it has a built-in drift wave instability for any finite value of $\alpha$, and because strong zonal flows are observed for moderate and large values of $\alpha$. However, the adiabatic limit, $\alpha\rightarrow \infty$, of the mHW model is problematic. Indeed, the mHW model does \textit{formally} converge to the modified Hasegawa-Mima model (mHM) \citep{dewar2007zonal} when $\alpha\rightarrow\infty$, as one would expect; however, for $\alpha$ large but finite, the dynamics of the mHW model can be significantly different from that of the mHM model, as we have shown in \citep{majda2018flux}. The bHW model, presented below, addresses this limitation.

\subsection{Formulation of the flux-balanced model}

The\emph{ flux-balanced Hasegawa-Wakatani model} (bHW) describes the time evolution of the balanced potential vorticity
$q=\nabla^{2}\varphi-\tilde{n}$ and particle density $n=\bar{n}+\tilde{n}$
according to \citep{majda2018flux} \addtocounter{equation}{0}\begin{subequations}\label{plasma_balanced}
\begin{eqnarray}
\frac{\partial q}{\partial t}+J\left(\varphi,q\right)-\kappa\frac{\partial\varphi}{\partial y} & = & D\Delta q,\label{eq:plasma_balc1}\\
\frac{\partial n}{\partial t}+J\left(\varphi,n\right)+\kappa\frac{\partial\varphi}{\partial y} & = & \alpha\left(\tilde{\varphi}-\tilde{n}\right)+D\Delta n,\label{eq:plasma_balc2}
\end{eqnarray}
\end{subequations}
We presented a detailed study of the properties of this model in \citep{majda2018flux}. Let us highlight its most important features. First, the equations have the expected invariance under Galilean transformations along the $y$-direction. Second, the equations describe the same linear drift wave instability as previous HW models. Third, the model converges to the modified Hasegawa-Mima (mHM) model in the limit of adiabatic electrons, as desired. In \citep{majda2018flux}, we give a rigorous proof as well as numerical evidence for this convergence. Formally, this can be seen as follows. In the limit $\alpha\rightarrow\infty$, Eq.(\ref{eq:plasma_balc2}) implies $\tilde{n}\rightarrow\tilde{\varphi}$, so that $q\rightarrow\nabla^{2}\varphi-\tilde{\varphi}$. Equation (\ref{eq:plasma_balc1}) then takes the form
\[
\frac{\partial q}{\partial t}+J\left(\varphi,q\right)-\kappa\frac{\partial\varphi}{\partial y}=0,\quad q=\nabla^{2}\varphi-\tilde{\varphi}.
\]
which is precisely the mHM equation \citep{hasegawa1978pseudo,dewar2007zonal}. The improved treatment of the parallel electron dynamics in the bHW model as compared to the mHW model, illustrated by the proper convergence of the model to the mHM model, leads to profound changes in the observed dynamics. As we will see below, the zonal structures in the bHW model are stronger, more persistent, and at the same time have larger variability.

We note that the system of equations (\ref{plasma_balanced}) is a simplified form of the bHW equations we studied in \citep{majda2018flux} in the sense that more generalized dissipation effects have been considered in \citep{majda2018flux}. We showed that these effects can play a significant role in the dynamics, particularly so for the simple model for Landau damping first introduced by \citet{wakatani1984collisional}. However, none of these dissipation effects can be rigorously derived from kinetic theory for the plasma regimes of magnetic confinement fusion interest, so for simplicity we will only consider the elementary forms of dissipation given on the right-hand sides of Eq. (\ref{plasma_balanced}).

\subsubsection{Spectral form of the bHW model}

For the numerical solution of the equations as well as the linear stability analysis, it is useful to rewrite the equations in Fourier space by considering the Fourier expansions of the electrostatic potential $\varphi$ and particle density $n$ assuming doubly periodic boundary condition
\[
\varphi=\sum_{k}\hat{\varphi}_{k}\exp\left(i\mathbf{k\cdot x}\right),\quad n=\sum_{k}\hat{n}_{k}\exp\left(i\mathbf{k\cdot x}\right),\quad\zeta=\sum_{k}-k^{2}\hat{\varphi}_{k}\exp\left(i\mathbf{k\cdot x}\right),
\]
with $\mathbf{k}=\left(k_{x},k_{y}\right)$ and $k^{2}=\left|\mathbf{k}\right|^{2}$.
The balanced potential vorticity $q=\zeta-\tilde{n}$ formulated in
the bHW model (\ref{plasma_balanced}) can be written as
\[
q=\sum_{k}\hat{q}_{k}\exp\left(i\mathbf{k\cdot x}\right),\quad\hat{q}_{k}=-k^{2}\hat{\varphi}_{k}-\left(1-\delta_{k_{y},0}\right)\hat{n}_{k},
\]
where we have removed the zonal mean $\overline{n}$ corresponding to the zonal modes $k_{y}=0$ by introducing the Kronecker delta operator. 

The bHW equations (\ref{plasma_balanced}) can be written for each spectral mode $\left(\hat{q}_{k},\hat{n}_{k}\right)$ as
\begin{equation}
\begin{aligned}\frac{d\hat{q}_{k}}{dt}+J\left(\varphi,q\right)_{k}-i\kappa k_{y}\hat{\varphi}_{k} & =-k^{2}D\hat{q}_{k},\\
\frac{d\hat{n}_{k}}{dt}+J\left(\varphi,n\right)_{k}+i\kappa k_{y}\hat{\varphi}_{k} & =\alpha\left(1-\delta_{k_{y},0}\right)\left(\hat{\varphi}_{k}-\hat{n}_{k}\right)-k^{2}D\hat{n}_{k}.
\end{aligned}
\label{eq:bhw_spectral}
\end{equation}
The quadratic terms $J\left(\varphi,q\right)_{k},J\left(\varphi,n\right)_{k}$ above include the triad interactions between modes at different scales. Once again, the zonal modes $k_{y}=0$ are treated separately by introducting the Kronecker delta $\delta_{k_{y},0}$ in the density equation, and the fact that it is present in the definition of $\hat{q}_{k}$.

For a general rectangular computational domain $\mathbf{x}=\left(x,y\right)\in\left[-L_{x}/2,L_{x}/2\right]\times\left[-L_{y}/2,L_{y}/2\right]$,
the wavenumber space is discrete, with wavenumber increments $\Delta k_{x}=2\pi/L_{x}$ and $\Delta k_{y}=2\pi/L_{y}$:
\[
k_{x}=\frac{2\pi}{L_{x}}n_{x},\;k_{y}=\frac{2\pi}{L_{y}}n_{y},\quad n_{x}=-\frac{N_{x}}{2}+1,\cdots,\frac{N_{x}}{2},\;n_{y}=-\frac{N_{y}}{2}+1,\cdots,\frac{N_{y}}{2}.
\]
A larger domain size in a given direction will thus create more intermediate wavenumbers in the corresponding direction in the spectral domain. For all the simulations we will show below, we used the same resolution $\frac{L_{x}}{N_{x}}=\frac{L_{y}}{N_{y}}$ in the $x$ and $y$ directions. The largest wavenumber for the $x$ and $y$ directions is therefore equal, $k_{x,\max}=k_{y,\max}=2\pi\frac{N_{x}}{L_{x}}=2\pi\frac{N_{y}}{L_{y}}$. 

\subsection{Linear stability analysis\label{sub:Stable-and-unstable}}

\begin{figure}
\includegraphics[scale=0.24]{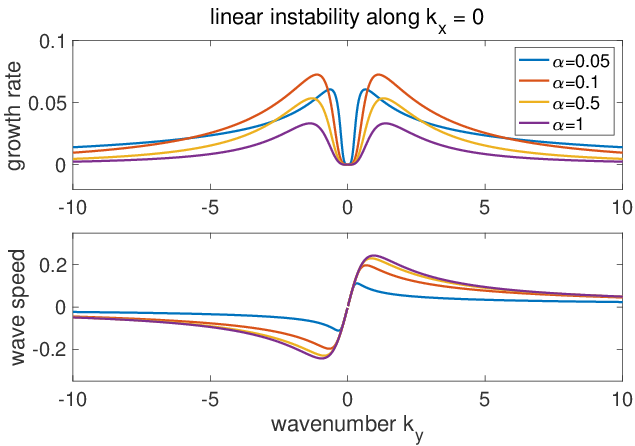}\includegraphics[scale=0.23]{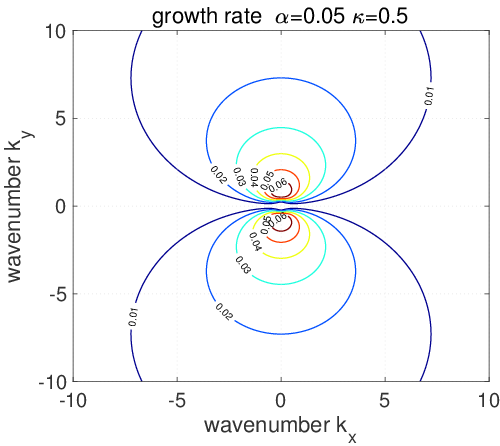}\includegraphics[scale=0.23]{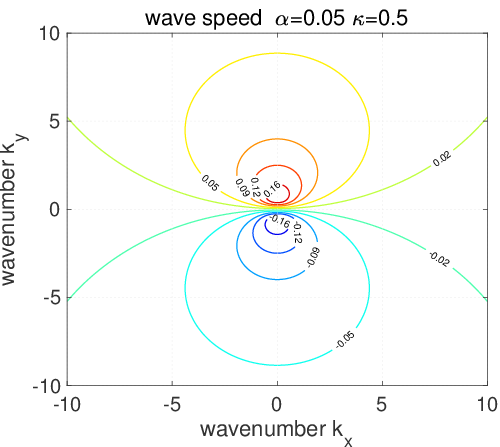}

\caption{Results of the linear instability analysis with $D=0$. The
left panel shows the linear growth rate and wave speed as a function of $k_{y}$ along the line $k_{x}=0$ for different values of $\alpha$. The next two plots
show contours of normalized growth rate and wave speed in $k_{x}-k_{y}$ space. If one increases the lengths $L_{x}$ and $L_{y}$ of the computational domain, more modes will be located in the region of strong instability.\label{fig:Linear-instability}}
\end{figure}

Before studying the nonlinear dynamics of the bHW model, it is valuable to establish the linear stability properties of the model. Here, we follow the standard procedure for linear stability analysis, in which we consider the growth of small perturbations corresponding to a single plane wave solution, $\varphi=\hat{\varphi}\exp\left(i\left(\mathbf{k\cdot x}-\omega t\right)\right)$, $n=\hat{n}\exp\left(i\left(\mathbf{k\cdot x}-\omega t\right)\right)$ with $k_{y}\neq0$. For zero background mean flow, the nonlinear interaction terms $J\left(\varphi,q\right)$ and $J\left(\varphi,n\right)$ vanish at the lowest order. Dropping the nonlinear interaction terms in
(\ref{eq:bhw_spectral}) gives us the linearized equations for the spectral coefficients $\left(\hat{\varphi},\hat{n}\right)$
\begin{equation}
\begin{aligned}i\omega k^{2}\hat{\varphi} & =\alpha\left(\hat{\varphi}-\hat{n}\right)+Dk^{4}\hat{\varphi},\\
-i\omega\hat{n} & =\alpha\left(\hat{\varphi}-\hat{n}\right)-i\kappa k_{y}\hat{\varphi}-Dk^{2}\hat{n}.
\end{aligned}
\label{eq:linear_dispersion}
\end{equation}
Now, we want to focus here on the stability boundary of the resistive drift wave instability. We thus keep the resistive dissipation term, which is the source of the drift wave instability, but set the other dissipation terms to zero by fixing $D=0$. Note that in the absence of other forms of dissipation than the resistive term, the linear stability analysis is identical in the bHW and mHW models. Non-trivial solutions to the equation (\ref{eq:linear_dispersion}) exist provided that the coefficient matrix has determinant zero. This condition gives us the resistive drift wave dispersion relation 
\begin{equation}\label{eq:drift_dispers}
\omega^2+ib(\omega-\omega_{*})=0
\end{equation}
where where the drift wave frequency $\omega_{*}$ and $b$ are given by
\begin{displaymath}
\omega_{*}(k)=\frac{\kappa k_{y}}{1+k^2}\qquad,\qquad b(k)=\alpha(1+k^{-2})
\end{displaymath}
Equation (\ref{eq:drift_dispers}) is readily solved, and we find the following expressions for the real and imaginary part of the frequency $\omega$:
\begin{equation}
\begin{aligned}
& \omega_{r}=\pm\frac{b}{2}\left(1+16\gamma^{2}\right)^{1/4}\cos\frac{\theta}{2},\;\omega_{i}=-\frac{b}{2}\pm\frac{b}{2}\left(1+16\gamma^{2}\right)^{1/4}\sin\frac{\theta}{2}, \\ 
& \theta=\mathrm{Arg}\left(-1+4\gamma i\right),\;\gamma=\frac{\omega_{*}}{b}.\label{eq:soln1}
\end{aligned}
\end{equation}
We showed in \citep{majda2018flux} that if $\alpha\ne0,\kappa\ne0$, all modes have $\omega_{i}>0$, i.e. all modes with $k_{y}\neq 0$ are unstable. Furthermore, the maximum growth rates are obtained for $k_{x}=0$. This can be seen for the particular case with $\alpha=0.05$ and $\kappa=0.5$ in the middle plot of Figure \ref{fig:Linear-instability}. As one would expect, there is no instability for the zonal modes, along the line $k_{y}=0$. In the first plot of Figure \ref{fig:Linear-instability}, we show the growth rate $\omega_{i}$ and the wave speed $\omega_{r}$ in terms of $k_{y}$ along the line $k_{x}=0$, for different values of the adiabaticity parameter $\alpha$. The growth rate increases from zero at $k_{y}=0$ to a maximum value before decreasing for large values of $k$. The position
$k_{\max}$ of the maximum growth wavenumber increases as the parameter value $\alpha$ increases, and reaches the asymptotic value $k_{max}=\sqrt{2}$ in the limit $\alpha\rightarrow\infty$. The maximum growth rate first increases with $\alpha$, but then decreases as $\alpha$ increases further. This is expected: HW models converge to HM limits when $\alpha\rightarrow\infty$, and HM models do not have an internal drift instability. In other words, when electrons are adiabatic, the drift waves propagate without growing or damping. Finally, for large values of $k$, i.e. at small scales, smaller values of $\alpha$ always correspond to larger positive growth rates, illustrating the generally more turbulent nature of the drift wave dynamics for small $\alpha$.

We conclude this subsection by observing that the aspect ratio of the computational domain does obviously not modify the linear stability analysis presented here. However, by increasing the size of the domain, one decreases the wavenumber increments, obtaining finer resolution in Fourier space. Looking at Figure \ref{fig:Linear-instability}, this means that there will be more strongly unstable modes. A larger domain will therefore usually correspond to stronger interactions between the unstable spectral modes.

\section{Flow Features in Drift Wave Turbulence and Zonal Jets Regimes\label{sec:Flow-Features}}

In tokamak edge turbulence experiments, one observes flow transitions from the drift wave turbulence dominated regime to the more regular strong zonal jet regime, in which zonal flows reduce cross field transport. Similar transitions are observed in both the mHW and the bHW models. In this section, we rely on direct numerical simulations to study the transition between the two flow regimes as one changes the adiabaticity parameter $\alpha$, and to compare the dynamics in the bHW and mHW models.

We will first consider a square geometry
$\left[-L/2,L/2\right]\times\left[-L/2,L/2\right]$ with $L=40$, a background density gradient with value $\kappa=0.5$, a cross-field diffusion coefficient $D=5\times10^{-4}$, and two typical values for the adiabaticity parameter $\alpha$: $\alpha=0.05$, corresponding to a strong drift wave turbulence regime, and $\alpha=0.5$, corresponding to a zonal jet dominated regime. The same set of parameters are used for both the mHW and bHW models. We use $N=256$ discretisation points, so that the smallest resolved scale is $\Delta x=\Delta y=40/256$. Our direct numerical simulations are based on a standard pseudo-spectral scheme, in which we calculate the nonlinear terms in real space instead of Fourier space. To stabilize the truncated numerical system, hyperviscosities, $\nu\Delta^{2s}q$ and $\nu\Delta^{2s}n$, are added in the potential vorticity and density equations with the viscosity strength $\nu=7\times10^{-21}$ and order $s=4$. A fourth-order explicit-implicit Runge-Kutta scheme is used for the time integration. The stiff hyperviscosity operator $v\Delta^{8}$ is integrated with an implicit scheme, while all the other terms are treated explicitly.

\subsection{Transition from drift wave turbulence to zonal flows}

In order to compare the major differences in the dynamics of the bHW model (\ref{plasma_balanced}) and the mHW model (\ref{plasma}), we start by rewriting both models in terms of the balanced potential vorticity $q=\nabla^{2}\varphi-\tilde{n}$: 
\begin{equation}
\begin{aligned}\mathrm{mHW:}\quad\;\:\frac{\partial}{\partial t}\left(\nabla^{2}\varphi-\tilde{n}\right)+J\left(\varphi,\nabla^{2}\varphi-\tilde{n}\right)+\partial_{x}\left(\overline{\tilde{u}\tilde{n}}\right)+\left(\partial_{x}\overline{n}-\kappa\right)\frac{\partial\tilde{\varphi}}{\partial y} & =D\Delta\left(\nabla^{2}\varphi-\tilde{n}\right),\\
\mathrm{bHW:}\qquad\frac{\partial}{\partial t}\left(\nabla^{2}\varphi-\tilde{n}\right)+J\left(\varphi,\nabla^{2}\varphi-\tilde{n}\right)\qquad\qquad\qquad\quad\quad-\kappa\frac{\partial\tilde{\varphi}}{\partial y} & =D\Delta\left(\nabla^{2}\varphi-\tilde{n}\right).
\end{aligned}
\label{eq:comp_BMHW}
\end{equation}
We observe that there are two additional terms in the mHW model, which are zonally
averaged quantities: the zonal density gradient term, $\overline{n}_{x}\frac{\partial\varphi}{\partial y}$, which effectively acts to modify the background density gradient profile $\kappa$; and the eddy flux, $\partial_{x}\left(\overline{\tilde{u}\tilde{n}}\right)$, corresponding to the zonal transport of particles in the radial direction. Both terms are zonal mean state feedbacks which are strongest at the largest scales, i.e. the scales of the zonal jet structures. We therefore expect the statistical properties of the bHW model and mHW model to differ significantly at the largest scales. The formation of the zonal structures is described by the equations for the zonal mean states $\overline{q}\left(x\right)=\partial_{x}^{2}\overline{\varphi}$
and $\overline{n}\left(x\right)$, which can be written as follows in the bHW model:
\begin{equation}
\begin{aligned}\partial_{t}\overline{q}+\partial_{x}\left(\overline{\tilde{u}\tilde{q}}\right) & =D\partial_{x}^{2}\overline{q},\\
\partial_{t}\overline{n}+\partial_{x}\left(\overline{\tilde{u}\tilde{n}}\right) & =D\partial_{x}^{2}\overline{n}.
\end{aligned}
\label{eq:plasma_mean-bhw}
\end{equation}
where $\tilde{q}=\nabla^{2}\tilde{\varphi}-\tilde{n}$ and $\tilde{u}=-\partial_{y}\tilde{\varphi}$. We note once more that like the mHW model but unlike the oHW model, the adiabaticity parameter $\alpha$ is absent from the zonal mean equations 

The feedback of the fluctuations to the zonal mean state is through the divergence of the nonlinear fluxes
\[
\partial_{x}\left(\overline{\tilde{u}\tilde{f}}\right)=\frac{1}{L_{y}}\int\left(\tilde{\varphi}_{x}\tilde{f}_{y}-\tilde{\varphi}_{y}\tilde{f}_{x}\right)dy,
\]
with $\tilde{f}=\tilde{q},\tilde{n}$. These flux terms play a central role in the energy transfer between the turbulent drift waves and the zonal structures. The generation of zonal flows in turn mediates turbulent transport by absorbing energy from the
turbulent drift waves. To see this, we consider the transition from a dominantly turbulent regime to a zonal flow dominated regime corresponding to an increase of the adiabaticity parameter $\alpha$, and look at the corresponding change in the total radial particle and vorticity fluxes, defined by 
\[
\Gamma_{n}=\frac{1}{L_{y}}\int_{0}^{L_{y}}\left(-\partial_{y}\tilde{\varphi}\right)\tilde{n}dy=\overline{\tilde{u}\tilde{n}},\quad\Gamma_{q}=\frac{1}{L_{y}}\int_{0}^{L_{y}}\left(-\partial_{y}\tilde{\varphi}\right)\tilde{q}dy=\overline{\tilde{u}\tilde{q}},
\]

\begin{figure}
\includegraphics[scale=0.32]{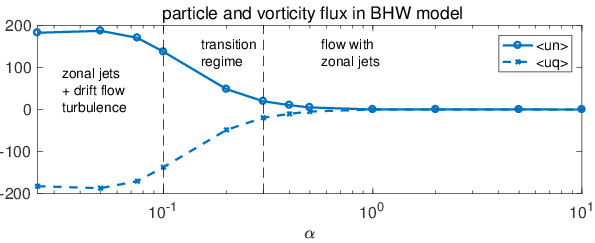}\includegraphics[scale=0.32]{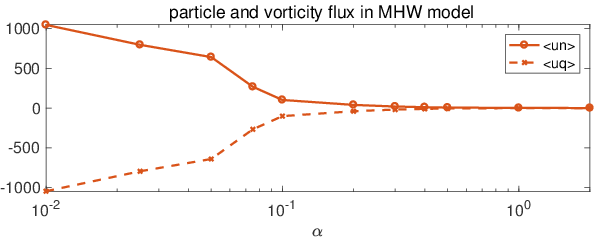}

\caption{Comparison of the total particle flux and the total vorticity flux as a function of $\alpha$, for simulations of the bHW model (left) and the mHW model (right) model. The other parameters are kept fixed as $\kappa=0.5$, $D=5\times10^{-4}$. Notice the much larger fluxes in the mHW model in the limit $\alpha\rightarrow0$, corresponding to a flow regime with many strong small-scale vortices. \label{fig:Total-statistical-energy} }
\end{figure}

\begin{figure}
\subfloat[turbulent regime $\alpha=0.05$]{\includegraphics[scale=0.21]{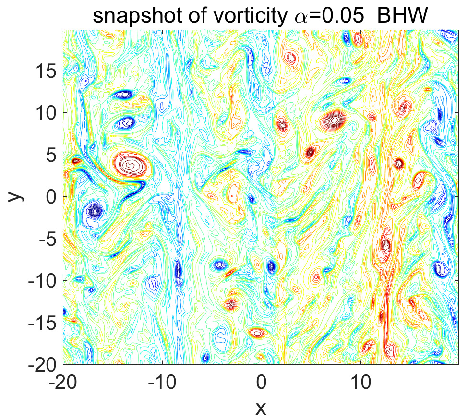}\includegraphics[scale=0.21]{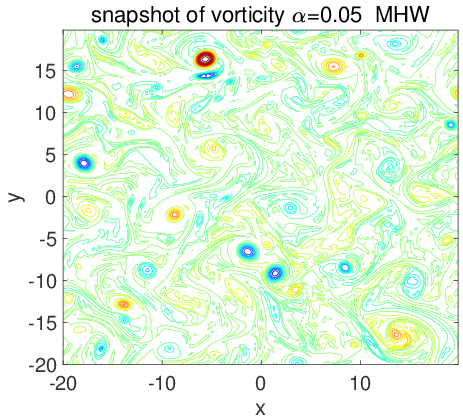}
}\subfloat[zonal flow regime $\alpha=0.5$]{\includegraphics[scale=0.21]{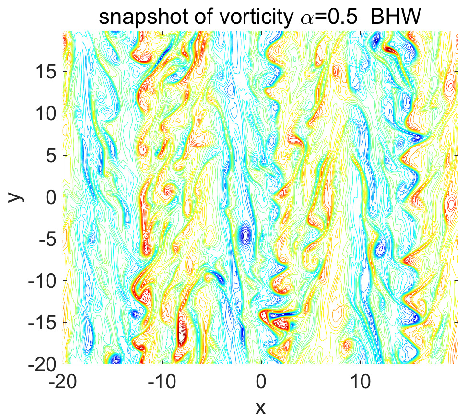}\includegraphics[scale=0.21]{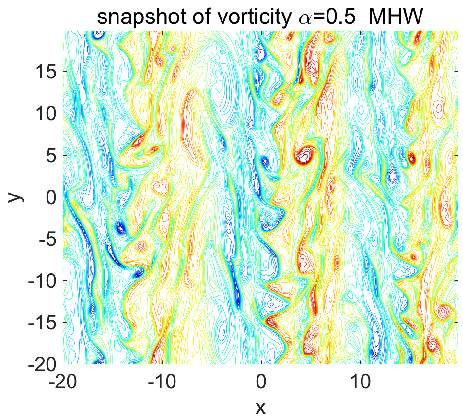}
}
\caption{Snapshots of the flow ion vorticity $\zeta=\nabla^{2}\varphi$ from
the bHW and mHW model simulations. Two typical regimes are compared:
a drift wave turbulence dominated regime $\alpha=0.05$ (upper panel),
and a zonal jet dominated regime $\alpha=0.5$ (lower panel).\label{fig:Snap_vort}}
\end{figure}

The results are shown in Figure \ref{fig:Total-statistical-energy}, in which $\Gamma_{n}$ and $\Gamma_{q}$ from the bHW model and the mHW model are plotted as a function of the parameter $\alpha$. Note that to generate this plot, we averaged the fluxes over time after the statistical equilibrium was reached. In both the bHW model and
the mHW model, the particle and vorticity fluxes decrease significantly when $\alpha$ increases, i.e. when the plasma resistivity decreases. Figure \ref{fig:Snap_vort}, which shows typical snapshots of the ion vorticity $\zeta=\nabla^{2}\varphi$ in the bHW model and in the mHW model for two representative regimes with
strong turbulence ($\alpha=0.05$) and strong zonal flows ($\alpha=0.5$), sheds light on the underlying physical process. For large values of $\alpha$, strong zonal jets are formed, which effectively block the radial transport of density and vorticity. In the strongly resistive limit, the zonal jets are weaker or nonexistent and the flow is significantly more turbulent, allowing strong particle fluxes.

Returning to Figure \ref{fig:Total-statistical-energy}, we observe a major difference between the bHW model and the mHW model. In the bHW model, the fluxes gradually increase as $\alpha$ decreases, but eventually saturate. In contrast, the fluxes never saturate as a function of $\alpha$ in the mHW model, and reach much higher values than in the bHW case. This too can be explained in the light of Figure \ref{fig:Snap_vort}. In the bHW model zonal structures persist in the small $\alpha$ regime, while in the mHW model the zonal jets disappear altogether, making way for fully homogeneous turbulence with many strong small vortices.

\begin{figure}
\begin{centering}
\includegraphics[scale=0.15]{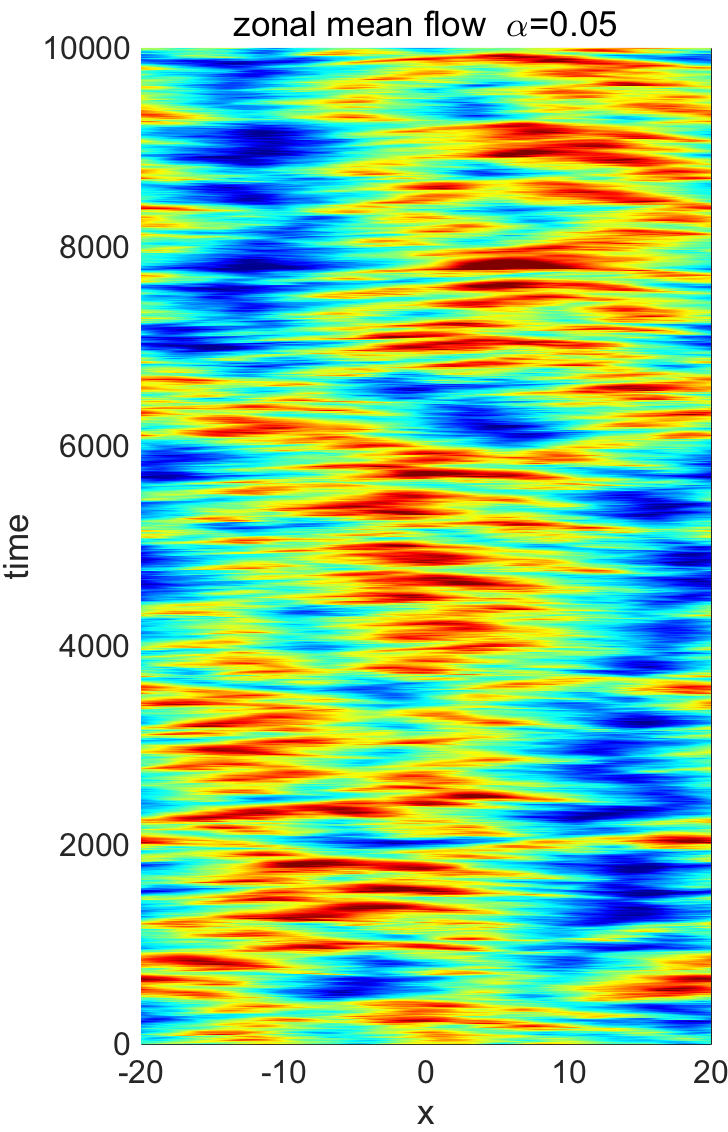}\includegraphics[scale=0.15]{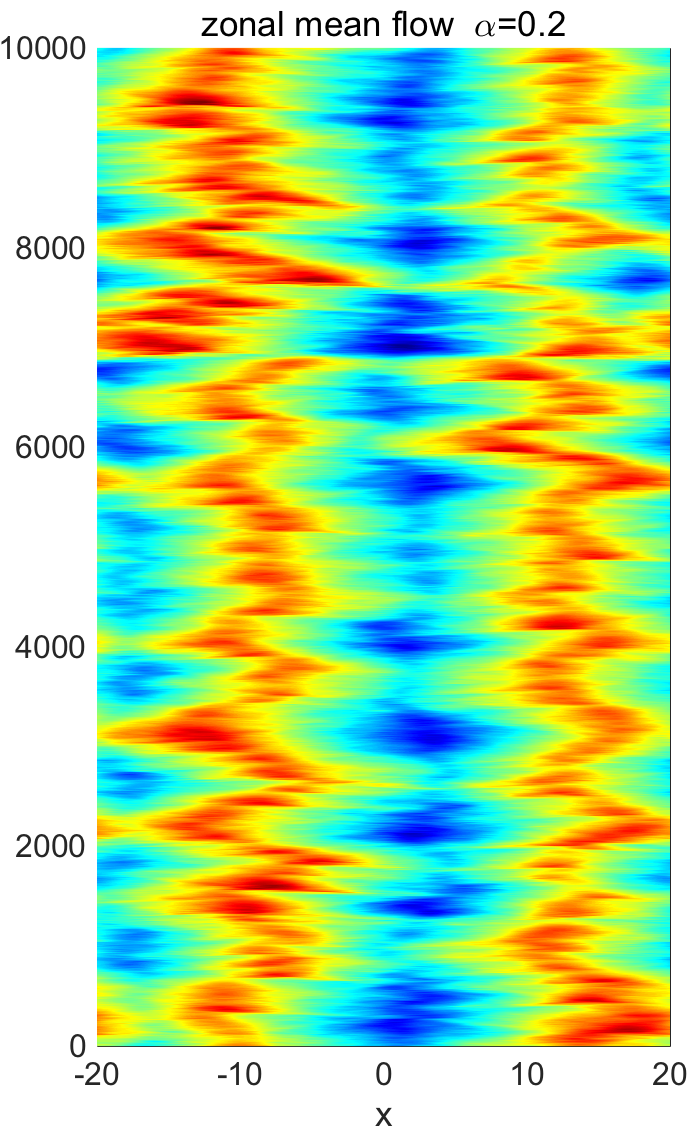}\includegraphics[scale=0.15]{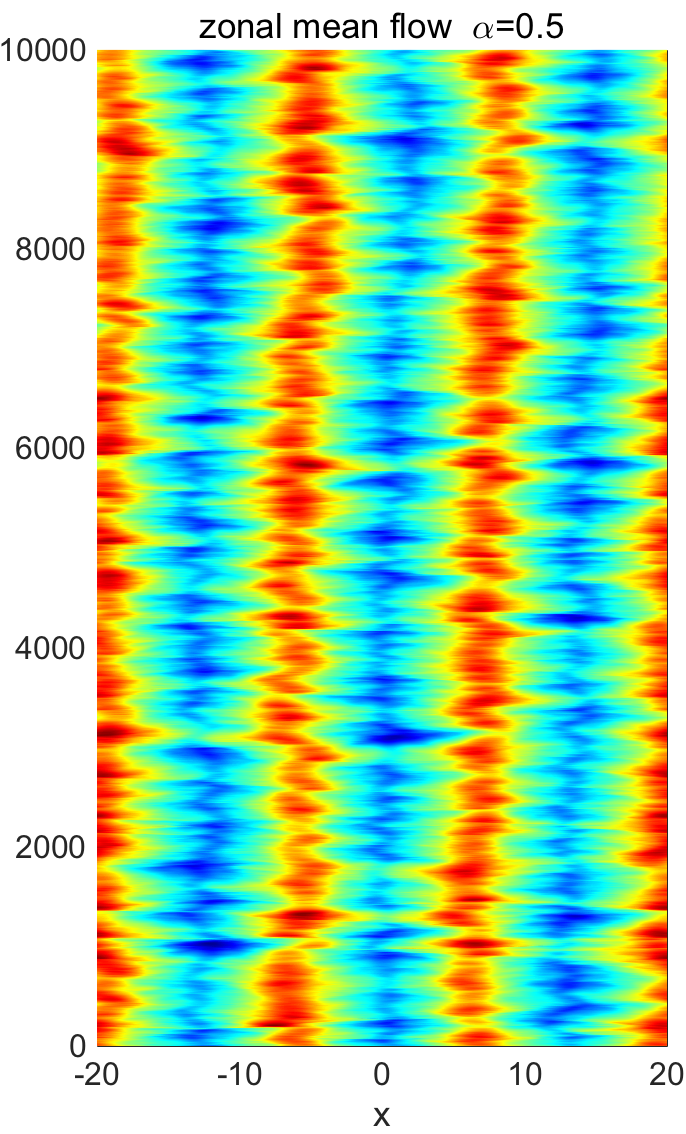}
\par\end{centering}

\caption{Time series of the zonally averaged mean flow $\overline{v}=\partial_{x}\overline{\varphi}$ for the typical, distinct regimes $\alpha=0.05,0.2,0.5$ in bHW simulations. The transition from rapidly shifting turbulent jets to strong zonal
structures can be observed as the value of $\alpha$ increases.\label{fig:Time-series-comp}}
\end{figure}

It is enlightening to look at the transition from the dominantly turbulent, small $\alpha$ regime to the zonally self-organized, large $\alpha$ regime from the point of view of the zonally averaged mean flow field $\bar{v}=\partial_{x}\bar{\varphi}$. This is what we do in Figure \ref{fig:Time-series-comp}, for the bHW model. As expected from our previous discussion, we always observe zonal jet structures, regardless of the value of $\alpha$. However, with smaller values of $\alpha$, the jets are fast shifting in space as they evolve in time. When $\alpha$ increases, two phenomena may be noticed: the jets become more stable, and a larger number of jets are generated. Still, we highlight the fact that the jets maintain a high level of spatial and temporal variability, even for large $\alpha$. This is a major difference with the mHW model, as we reported in \citep{majda2018flux} and also show in Figure \ref{fig:Time-series-v4}.

\subsection{Statistical spectra of the energy and the particle flux}\label{sec:spectra_square}

For a more detailed characterisation of the turburlence dominated regime and the zonal flow dominated regime, we focus on the statistical properties of the model state variables and compare the statistical spectra in both the bHW and mHW models for these two distinct regimes. Assuming ergodicity in the system, the statistics are calculated by averaging along the solution trajectory once the system has reached the statistically stationary state. We decompose any state variable of interest $f$ into its time averaged mean $\left\langle f\right\rangle _{\mathrm{eq}}$ and the fluctuations in time about the mean state $f^{\prime}=f-\left\langle f\right\rangle _{\mathrm{eq}}$. We concentrate here on the mean and on the variance of each mode, which tells us about the statistical variability at each scale. They are given by 
\[
\left\langle \hat{f}_{k}\right\rangle _{\mathrm{eq}}=\frac{1}{T}\int_{t_{0}}^{T}\hat{f}_{k}\left(t\right)dt,\quad\left\langle \left|\hat{f}_{k}^{\prime}\right|^{2}\right\rangle _{\mathrm{eq}}=\frac{1}{T}\int_{t_{0}}^{T}\left|\hat{f}_{k}\left(t\right)-\left\langle \hat{f}_{k}\right\rangle _{\mathrm{eq}}\right|^{2}dt,
\]
where $\hat{f}_{k}$ is the spectral mode of $f$ and the time average starts at the time $t_{0}$ when the statistical equilibrium state is reached.

\begin{figure}
\subfloat{\includegraphics[scale=0.22]{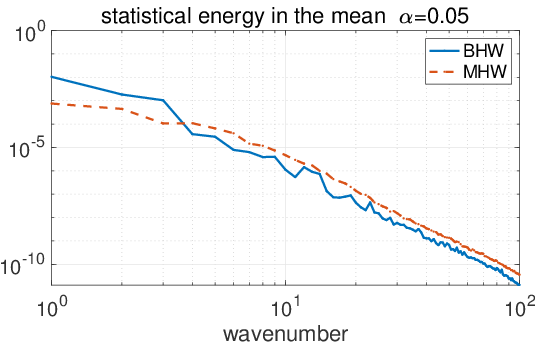}\includegraphics[scale=0.22]{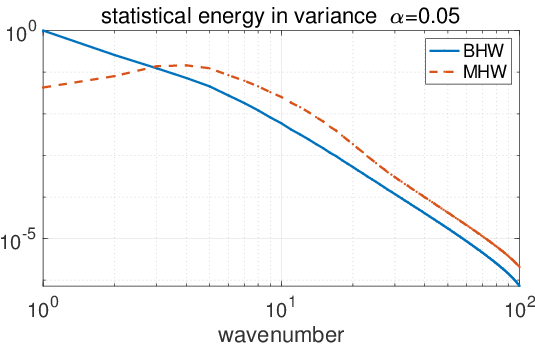}\includegraphics[scale=0.22]{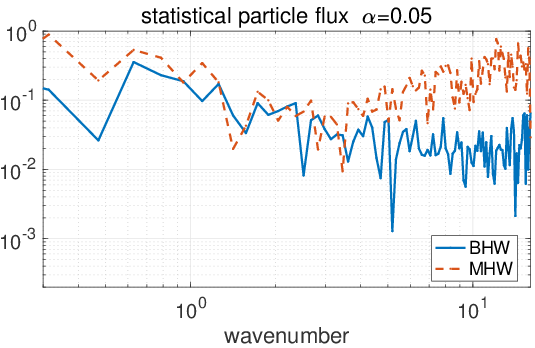}}

\subfloat{\includegraphics[scale=0.22]{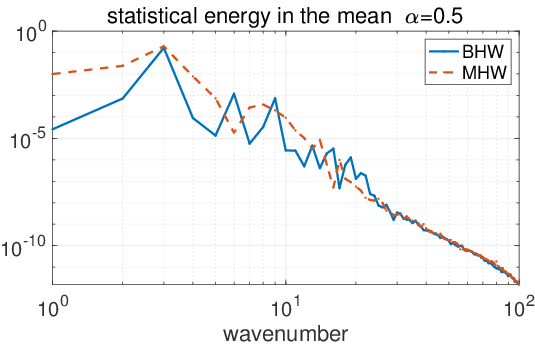}\includegraphics[scale=0.22]{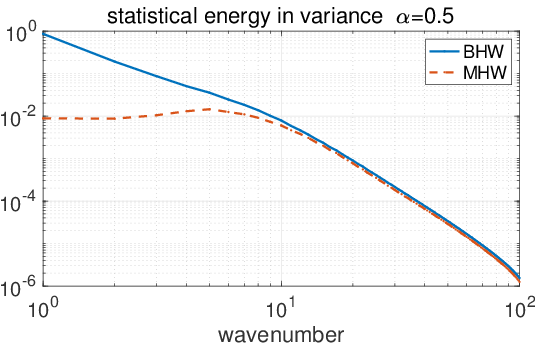}\includegraphics[scale=0.22]{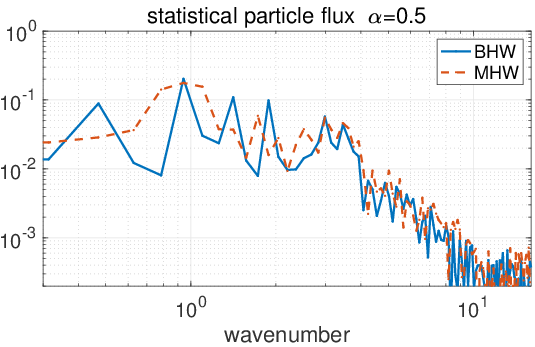}}

\caption{Spectra of the kinetic energy in the mean $|\langle k\hat{\varphi}_{k}\rangle_{\mathrm{eq}}|^2$ (left column) and in the variance $\langle|k\hat{\varphi}_{k}'|^2\rangle_{\mathrm{eq}}$ (center column), and spectra of the zonal particle flux $\left(\overline{\tilde{u}\tilde{n}}\right)_{k}$ (right), for both the bHW and the mHW models. The plots in the first row were obtained for $\alpha=0.05$, which corresponds to a turbulence-dominated regime, and the plots in the second row were obtained for $\alpha=0.5$, which corresponds to a zonal jet dominated regime. The bHW
model has more energy in the variance for the large scale modes. \label{fig:Spectra-of-ene}}
\end{figure}

Our results are given in Figure \ref{fig:Spectra-of-ene}. In the first two columns, we plot
the spectra for the kinetic energy in the mean $|\langle k\hat{\varphi}_{k}\rangle_{\mathrm{eq}}|^2$ and in the variance $\langle|k\hat{\varphi}_{k}'|^2\rangle_{\mathrm{eq}}$ for both the bHW and the mHW models, for the usual values $\alpha=0.05$, corresponding to the strongly turbulent case, in the top row, and $\alpha=0.5$ corresponding to strong jets, in the bottom row. For these spectra, the wavenumbers were rescaled into integer numbers. The results confirm our visual analysis from the previous subsection. At high resistivity, the kinetic energy in the mean is larger in the bHW model at large scales, indicating the presence of zonal structures which are absent in this regime in the mHW model. At the same time, the mHW model has larger energy in the variance at small scales, which indicates a more strongly turbulent regime with many small scale vortices. In the low resistivity regime, the kinetic energy in the mean has a dominant peak at $k=3$ in both models, which is a clear signature of the presence of three strong zonal jets. The bHW model has much more energy in the variance at large scales, which demonstrates that the jets in that model have more variability than in the mHW model.  

In the third column of Figure \ref{fig:Spectra-of-ene}, we plot spectra of the zonal particle flux $\left(\overline{\tilde{u}\tilde{n}}\right)_{k}$. This quantity is particularly interesting because it plays an important role in the equation for the potential vorticity in the mHW model, as shown in Eq. (\ref{eq:comp_BMHW}), and it is absent from the analogous equation in the bHW model. It also represents the feedback of the non-zonal modes to the zonal mean state in both models, as can be seen in Eq. (\ref{eq:plasma_mean-bhw}). For $\alpha=0.05$, one finds a much larger zonal particle flux in the mHW model, particularly at small scales. Energy is therefore injected into small-scale vortices, which leads to a full breakdown to homogeneous turbulence in the mHW model. On the other hand, for $\alpha=0.5$, the particle flux is particularly strong at
intermediate scales, around $k\sim1$. This is what leads to the generation of the multiple zonal jets observed in this regime in both models.

\subsection{Probability density functions of the state variables\label{sub:Probability-density-functions}}

\begin{figure}
\subfloat[Turbulence dominated regime $\alpha=0.05,\kappa=0.5$]{\includegraphics[scale=0.18]{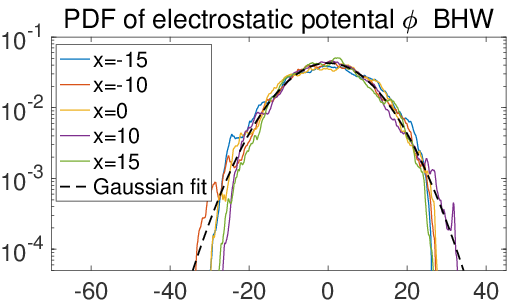}\includegraphics[scale=0.18]{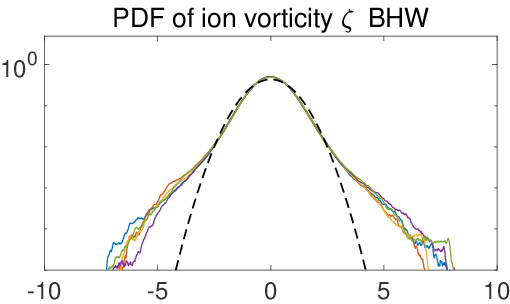}\includegraphics[scale=0.18]{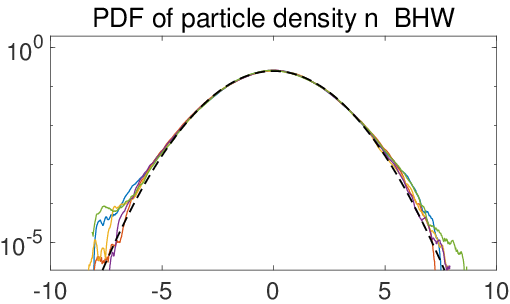}\includegraphics[scale=0.18]{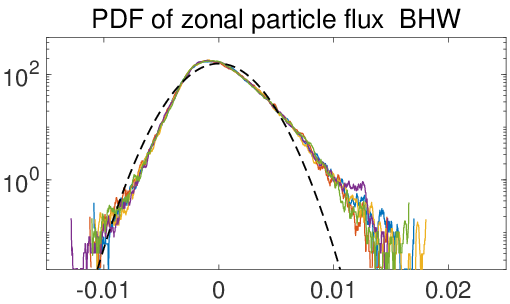}

}

\subfloat[Zonal flow dominated regime $\alpha=0.5,\kappa=0.5$]{\includegraphics[scale=0.18]{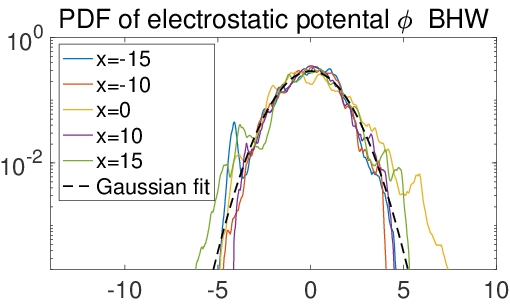}\includegraphics[scale=0.18]{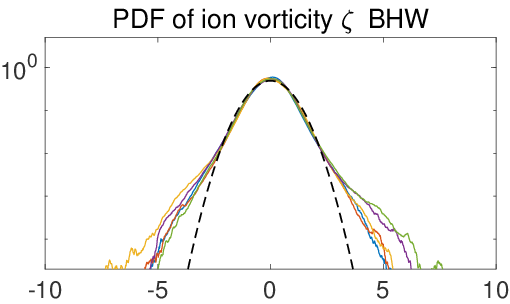}\includegraphics[scale=0.18]{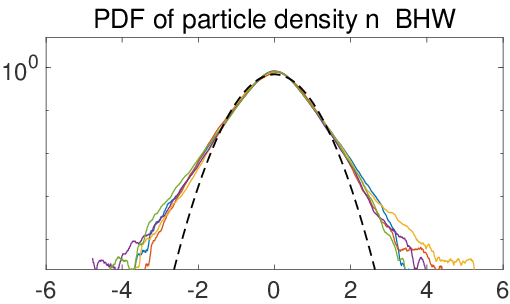}\includegraphics[scale=0.18]{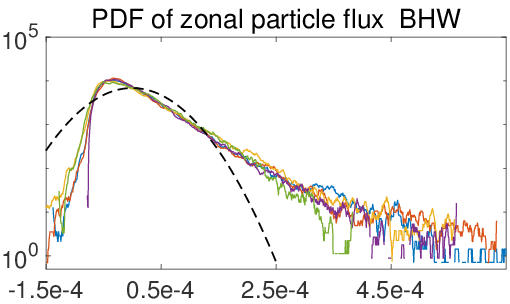}

}

\caption{Probability distribution functions of the electrostatic potential
$\varphi$, ion vorticity $\zeta$, particle density $n$, and zonal
particle flux $\overline{\tilde{u}\tilde{n}}$ in the bHW model, sampled at the five radial
locations given by $x=-15,-10,0,10,15$. The PDFs are shifted
to be centered at zero, and the mean values are listed in the legend.
\label{fig:Probability-distribution-function1}}
\end{figure}

In this last subsection, we compare the probability distribution functions
(PDFs) of the state variables in the two typical regimes, $\alpha=0.05$ and $\alpha=0.5$. Instead of focusing on the first two statistical moments, as we did previously, we are interested here in the non-Gaussian features of the PDFs, which are usually key signatures of turbulent systems and play an important role in determining the final balanced energy mechanisms of the system. In Figure \ref{fig:Probability-distribution-function1}, we show the probability density functions for the bHW model sampled for $\varphi,\zeta,n,\overline{\tilde{u}\tilde{n}}$ and for the two typical test regimes
$\alpha=0.05$ and $\alpha=0.5$ at five different $x$-locations in the computational domain $[-L/2,L/2]=[-20,20]$: $x=-15,-10,0,10,15$. We compute these PDFs by collecting histograms from the time-series of the stationary solutions. The electrostatic potential gives more emphasis to the larger scales, while the ion vorticity $\zeta = \nabla^{2}\varphi$ accentuates the smaller scale modes due to the additional factor $k^2$ for each spectral mode. As expected, the variances are much smaller for the zonal jet dominated regime, i.e. for $\alpha=0.5$, than for the turbulence-dominated regime, for $\alpha=0.05$. The distributions for the electrostatic potential $\varphi$ are approximately Gaussian. The ion vorticity $\zeta$, on the other hand, displays fat tails in the two regimes. The fatter
tails are the result of the intermittent fluctuating vortices appearing occasionally at small scales as the flow evolves. We distinguish two different behaviours for the density depending on the $\alpha$ regime: the density is almost Gaussian in the turbulent regime $\alpha=0.05$, but develops non-Gaussian fat tails in the zonal jet dominated regime $\alpha=0.5$. Finally, the zonal particle flux $\overline{\tilde{u}\tilde{n}}$ always has highly skewed distributions, corresponding to the net transport of particles in the direction opposite to the background density gradient. For $\alpha=0.5$, the strong zonal jets strongly reduce the flux $\overline{\tilde{u}\tilde{n}}$ as compared to the $\alpha=0.05$ case, as we have already highlighted. The flux in the $\alpha=0.5$ regime is much more skewed than in the other regime, with a fat tail corresponding to strong intermittency. These intermittent PDFs are typical in many turbulent flow fields and worth further investigation \citep{qi2016predicting,qi2018predicting}. They suggest that model reduction strategies based on standard quasilinear Gaussian approaches are likely to be inaccurate.

In Table \ref{tab:Basic-statistics}, we provide a compact summary of these results, listing the mean $\left\langle f\right\rangle _{\mathrm{eq}}$, variance $\left\langle |f^{\prime}|^{2}\right\rangle _{\mathrm{eq}}$, skewness $\left\langle |f^{\prime}|^{3}\right\rangle _{\mathrm{eq}}$, and kurtosis $\left\langle |f^{\prime}|^{4}\right\rangle _{\mathrm{eq}}$ (with the value 3 for Gaussian distributions) for the variables $f=\varphi,\zeta,n$, as well as the zonal particle flux $f=\overline{\tilde{u}\tilde{n}}$. For comparison, we computed the same quantities for the mHW model, and entered them in the same table. 

\begin{table}
\subfloat{{\footnotesize{}}%
\begin{tabular}{cccccccccccc}
\toprule 
{\footnotesize{}bHW} & \multicolumn{4}{c}{{\footnotesize{}$\alpha=0.05$}} &  &  & \multicolumn{4}{c}{{\footnotesize{}$\alpha=0.5$}} & \tabularnewline
\midrule 
 & {\footnotesize{}$\varphi$} & {\footnotesize{}$\zeta$} & {\footnotesize{}$n$} & {\footnotesize{}$\overline{\tilde{u}\tilde{n}}$} &  &  & {\footnotesize{}$\varphi$} & {\footnotesize{}$\zeta$} & {\footnotesize{}$n$} & {\footnotesize{}$\overline{\tilde{u}\tilde{n}}$} & \tabularnewline
\midrule
\midrule 
{\footnotesize{}mean} & {\footnotesize{}0.22} & {\footnotesize{}0.012} & {\footnotesize{}-0.012} & {\footnotesize{}0.0027} &  &  & {\footnotesize{}2.02} & {\footnotesize{}-0.27} & {\footnotesize{}-0.17} & {\footnotesize{}$0.78\!\!\times\!\!10^{-4}$} & \tabularnewline
\midrule 
{\footnotesize{}variance} & {\footnotesize{}86.92} & {\footnotesize{}0.82} & {\footnotesize{}2.49} & {\footnotesize{}$0.24\!\!\times\!\!10^{-3}$} &  &  & {\footnotesize{}1.88} & {\footnotesize{}0.11} & {\footnotesize{}0.057} & {\footnotesize{}$0.90\!\!\times\!\!10^{-6}$} & \tabularnewline
\midrule 
{\footnotesize{}skewness} & {\footnotesize{}-0.060} & {\footnotesize{}-0.014} & {\footnotesize{}-0.017} & {\footnotesize{}0.73} &  &  & {\footnotesize{}0.042} & {\footnotesize{}-0.045} & {\footnotesize{}0.25} & {\footnotesize{}2.19} & \tabularnewline
\midrule 
{\footnotesize{}kurtosis} & {\footnotesize{}2.54} & {\footnotesize{}5.40} & {\footnotesize{}3.11} & {\footnotesize{}4.51} &  &  & {\footnotesize{}3.55} & {\footnotesize{}6.53} & {\footnotesize{}7.79} & {\footnotesize{}12.07} & \tabularnewline
\bottomrule
\end{tabular}{\footnotesize \par}

}

\subfloat{{\footnotesize{}}%
\begin{tabular}{cccccccccccc}
\toprule 
{\footnotesize{}mHW} & \multicolumn{4}{c}{{\footnotesize{}$\alpha=0.05$}} &  &  & \multicolumn{4}{c}{{\footnotesize{}$\alpha=0.5$}} & \tabularnewline
\midrule 
 & {\footnotesize{}$\varphi$} & {\footnotesize{}$\zeta$} & {\footnotesize{}$n$} & {\footnotesize{}$\overline{\tilde{u}\tilde{n}}$} &  &  & {\footnotesize{}$\varphi$} & {\footnotesize{}$\zeta$} & {\footnotesize{}$n$} & {\footnotesize{}$\overline{\tilde{u}\tilde{n}}$} & \tabularnewline
\midrule
\midrule 
{\footnotesize{}mean} & {\footnotesize{}0.021} & {\footnotesize{}0.0083} & {\footnotesize{}-0.030} & {\footnotesize{}0.0122} &  &  & {\footnotesize{}2.35} & {\footnotesize{}-0.30} & {\footnotesize{}-0.16} & {\footnotesize{}$0.93\!\!\times\!\!10^{-4}$} & \tabularnewline
\midrule 
{\footnotesize{}variance} & {\footnotesize{}7.86} & {\footnotesize{}2.29} & {\footnotesize{}9.85} & {\footnotesize{}$6.40\!\!\times\!\!10^{-3}$} &  &  & {\footnotesize{}0.17} & {\footnotesize{}0.12} & {\footnotesize{}0.075} & {\footnotesize{}$0.93\!\!\times\!\!10^{-6}$} & \tabularnewline
\midrule 
{\footnotesize{}skewness} & {\footnotesize{}-0.068} & {\footnotesize{}-0.019} & {\footnotesize{}0.027} & {\footnotesize{}0.67} &  &  & {\footnotesize{}0.35} & {\footnotesize{}-0.52} & {\footnotesize{}0.85} & {\footnotesize{}2.09} & \tabularnewline
\midrule 
{\footnotesize{}kurtosis} & {\footnotesize{}3.38} & {\footnotesize{}17.21} & {\footnotesize{}3.17} & {\footnotesize{}4.36} &  &  & {\footnotesize{}2.45} & {\footnotesize{}5.76} & {\footnotesize{}5.50} & {\footnotesize{}14.92} & \tabularnewline
\bottomrule
\end{tabular}}

\caption{Basic statistics of the electrostatic potential $\varphi$, flow ion
vorticity $\zeta$, particle density fluctuation $n$, and zonal particle
flux $\overline{\tilde{u}\tilde{n}}$, at the sampling positions $x=0$.\label{tab:Basic-statistics}}

\end{table}

\section{Simulations of the bHW Model with Different Aspect Ratios\label{sec:Aspect-ratios}}

In this section, we consider the effect of the aspect ratio of the simulation domain on the flow dynamics described by the bHW model. Specifically, we consider a case with a domain extended in the $x$-direction, $\left[-5L/2,5L/2\right]\times\left[-L/2,L/2\right]$, a case with a domain extended in the $y$-direction, $\left[-L/2,L/2\right]\times\left[-5L/2,5L/2\right]$, and compare these two cases with the standard situation with a square domain $\left[-L/2,L/2\right]\times\left[-L/2,L/2\right]$ as we had in Section \ref{sec:Flow-Features}. Returning to Figure \ref{fig:Illustration} we expect the case with an extended domain in the $x$ direction to allow for a larger number of zonal jets which will interact with each other, while the second case should correspond to a situation with fewer and more extended jets. Both cases may be viewed as extreme cases which can qualitatively capture some typical features of realistic plasma flows.

\subsection{Typical plasma flow structures with different aspect ratios}

\subsubsection{Domain extended in the $x$-direction with aspect ratio $5:1$}

\begin{figure}
\includegraphics[scale=0.27]{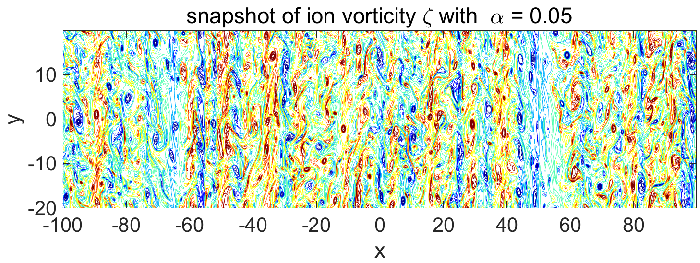}\includegraphics[scale=0.27]{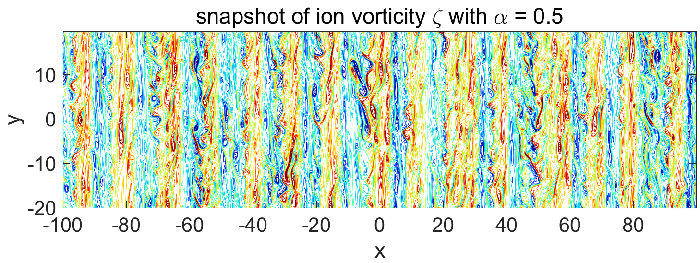}

\includegraphics[scale=0.27]{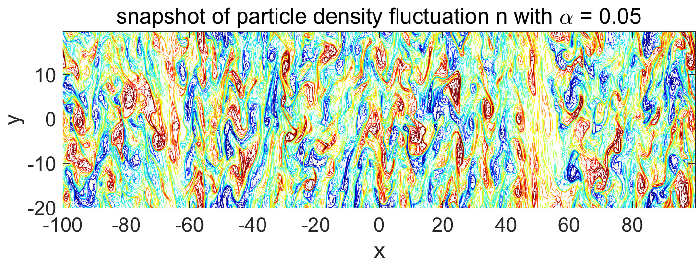}\includegraphics[scale=0.27]{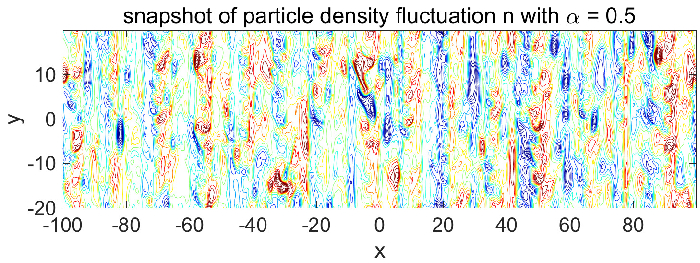}

\caption{Snapshots of the vorticity field $\zeta=\nabla^{2}\varphi$
and the density fluctuation $n$ from simulations of the bHW model, for an aspect ratio 5:1 in the high resistivity regime $\alpha=0.05,\kappa=0.5$ and the low resistivity
regime $\alpha=0.5,\kappa=0.5$.\label{fig:snapshots-of-flow1}}
\end{figure}

Let us start with simulations for an aspect ratio $L_{x}/L_{y}=5$. This is a wide and short simulation domain, in which a larger number of zonal jets will be obtained, allowing us to observe the complicated interactions between multiple jets. As we did before, we fix $\kappa=0.5$ and focus on the two characteristic regimes corresponding to high collisionality and low collisionality, $\alpha=0.05$ and $\alpha=0.5$ respectively.
The results are in agreement with the results we obtained for a square computational domain. For larger values of $\alpha$, stronger zonal jets develop, and for small value of $\alpha$ the flow is more turbulent with more interacting small scale vortices. This can be seen in Figure \ref{fig:snapshots-of-flow1}, which shows snapshots of the
vorticity $\zeta$ and density fluctuation $n$ for the two different values of $\alpha$.

Figure \ref{fig:snapshots-of-flow1} also illustrates the important fact that the results we get in this case could not be obtained by a mere five-fold duplication of the results we had for a square domain. Stronger turbulence is found in both regimes as compared to the situation with a square domain. Both regimes display multiple zonal jets with many small-scale fluctuation vortices, and the zonal jets have an interesting tendency to further organize into groups of larger scales in their evolution, as we will demonstrate more convincingly when we analyze the time-series of zonal mean flow. 

\subsubsection{Domain extended in the $y$-direction with aspect ratio $5:1$}

\begin{figure}
\subfloat[$\alpha=0.05,\kappa=0.5$]{\includegraphics[scale=0.28]{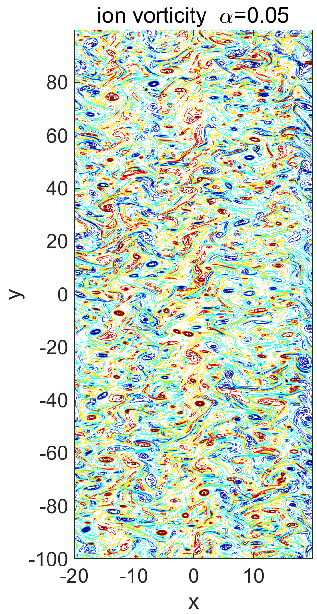}\includegraphics[scale=0.28]{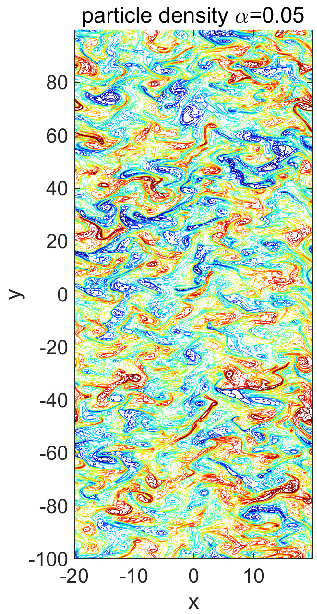}

}\subfloat[$\alpha=0.5,\kappa=0.5$]{\includegraphics[scale=0.28]{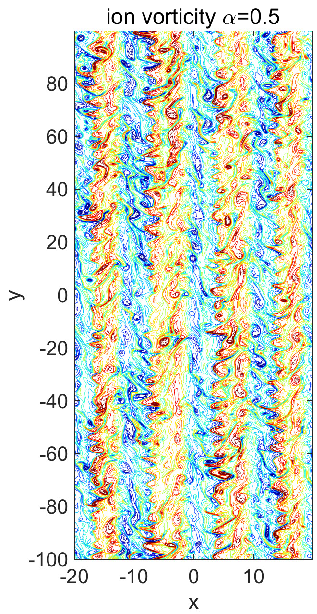}\includegraphics[scale=0.28]{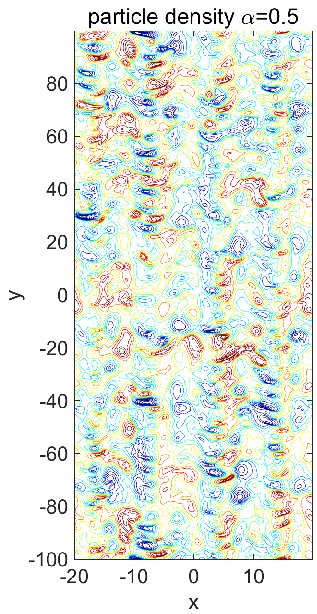}

}

\caption{Snapshots of the vorticity field $\zeta=\nabla^{2}\varphi$
and the density fluctuation $n$ from simulations of the bHW model, for an aspect ratio 1:5 in the high resistivity regime $\alpha=0.05,\kappa=0.5$ and the low resistivity
regime $\alpha=0.5,\kappa=0.5$.\label{fig:snapshots-of-flow3}}
\end{figure}

We now repeat the previous analysis, but this time for a computational domain extended in the $y$ direction, with aspect ratio $L_{x}/L_{y}=1/5$. Figure \ref{fig:snapshots-of-flow3} shows the vorticity $\zeta$ and density fluctuation $n$ for this case. Once more, zonal structures can be seen in both the collisional regime and the collisionless regime. Comparing these results with the case with a square domain, we note that many more energetic small scale structures are excited. In the strong resistivity case, $\alpha=0.05$, many small vortices are generated on top of the three dominant zonal jets. In the low resistivity case, $\alpha=0.5$, the longer zonal jets are less steady, and a 4-jet zonal structure develops, compared with the 3 jets seen for the square domain.

\subsubsection{Evolution of the zonal jets for different aspect ratios}

\begin{figure}
\subfloat[bHW $\alpha=0.05,\kappa=0.5$]{\includegraphics[scale=0.14]{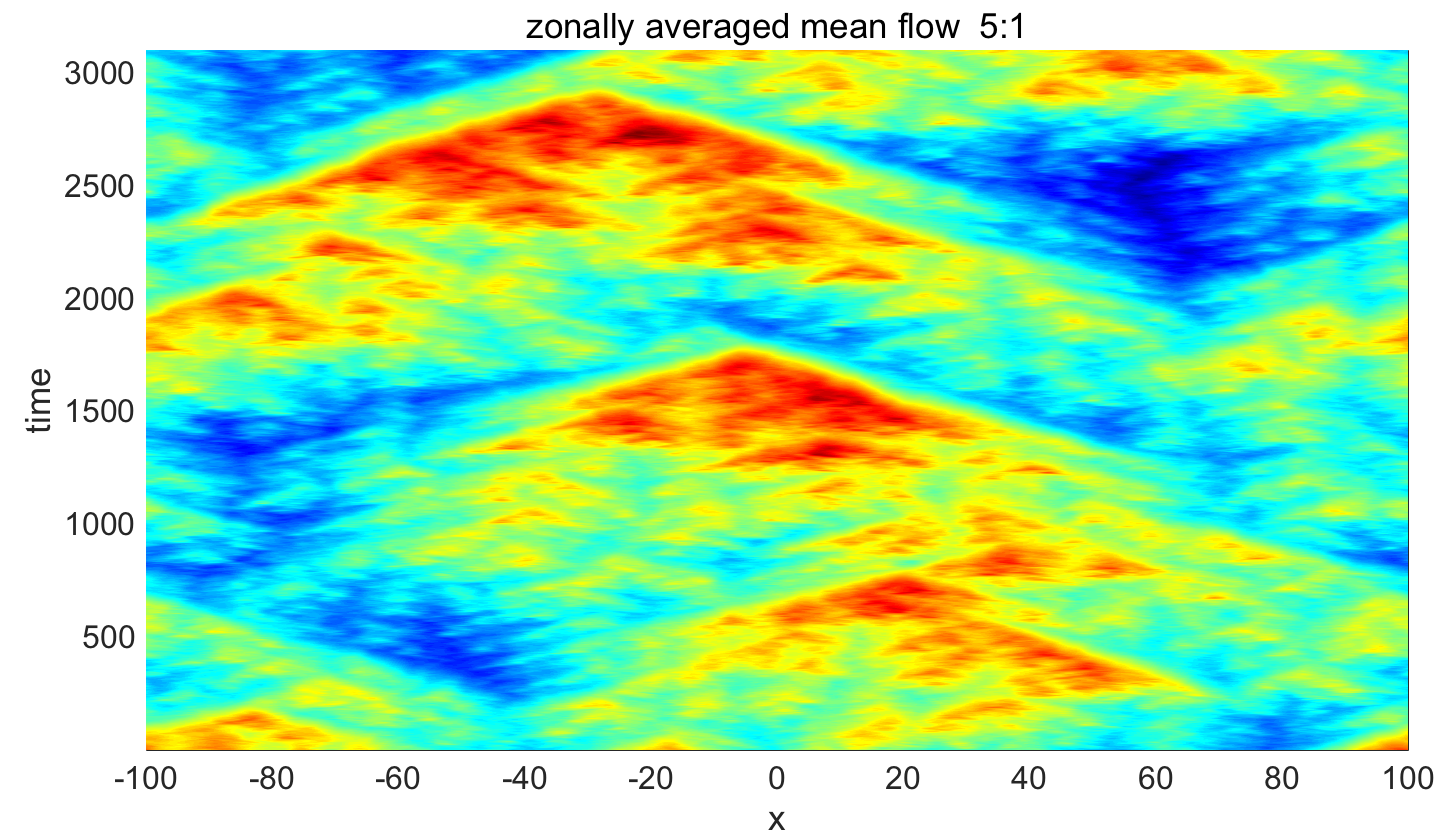}\includegraphics[scale=0.14]{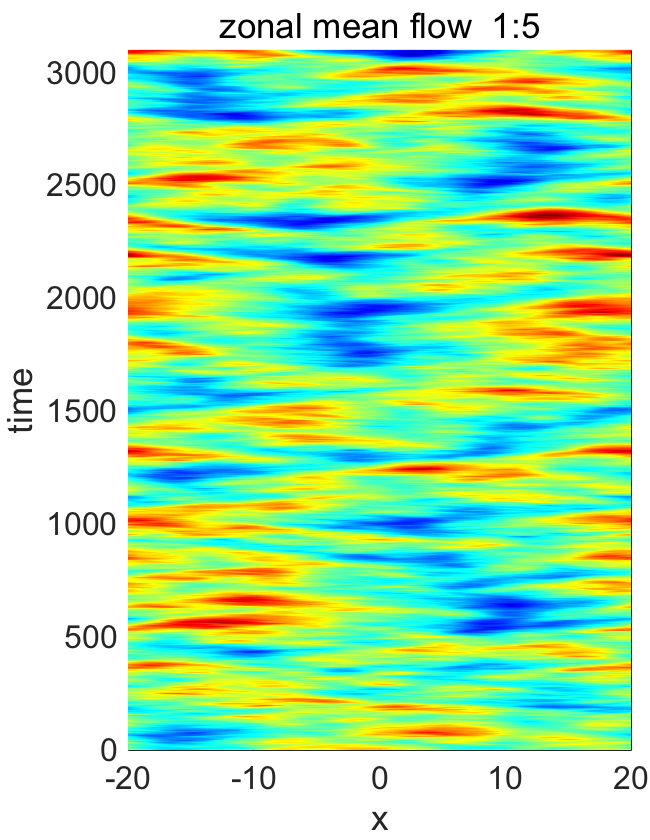}\includegraphics[scale=0.14]{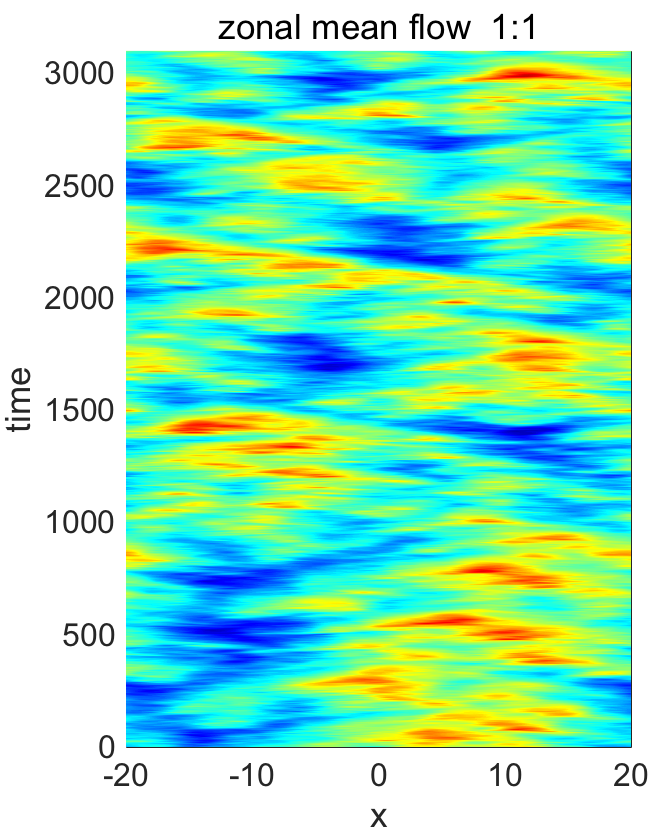}
}

\subfloat[bHW $\alpha=0.5,\kappa=0.5$]{\includegraphics[scale=0.14]{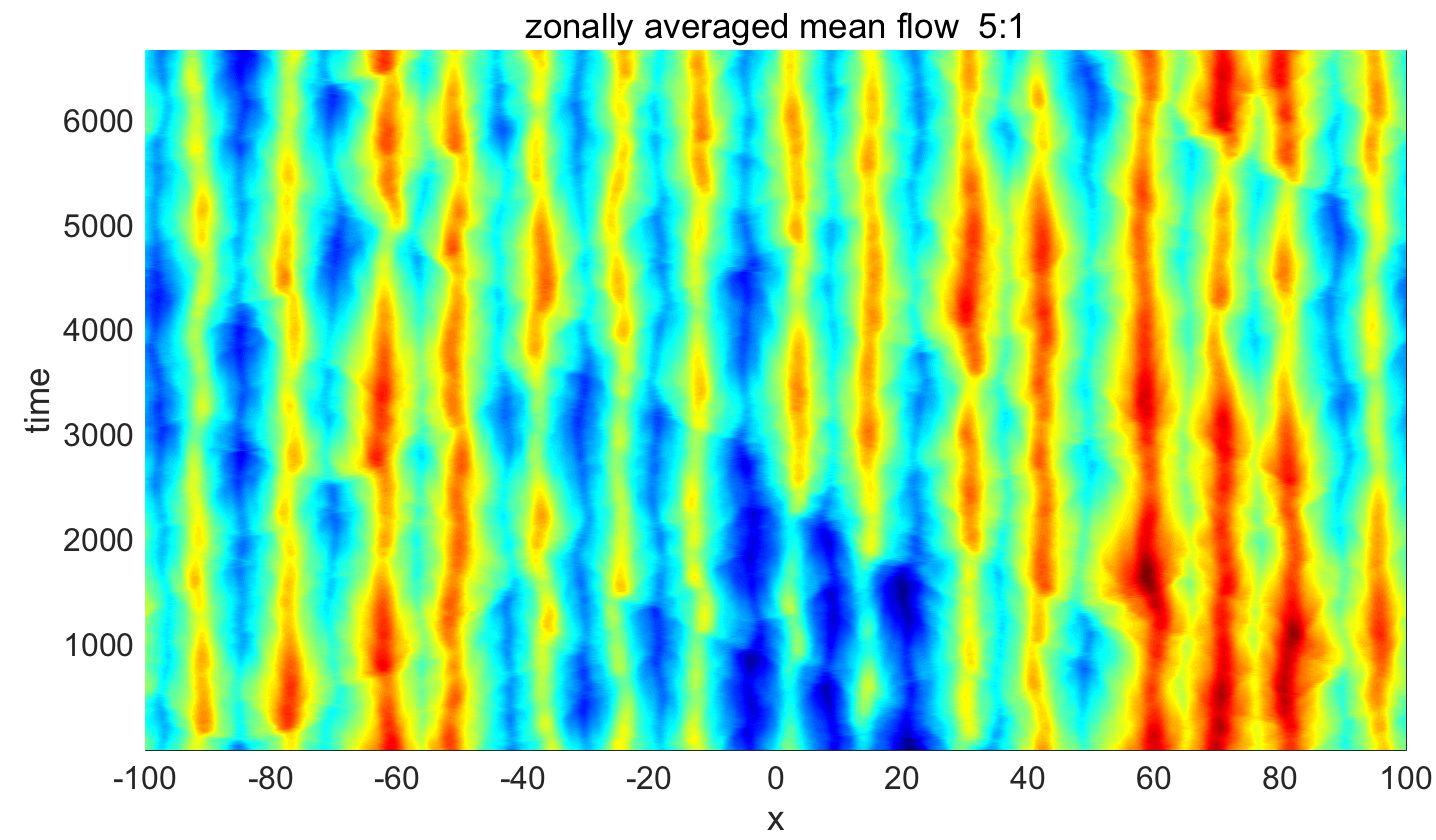}\includegraphics[scale=0.14]{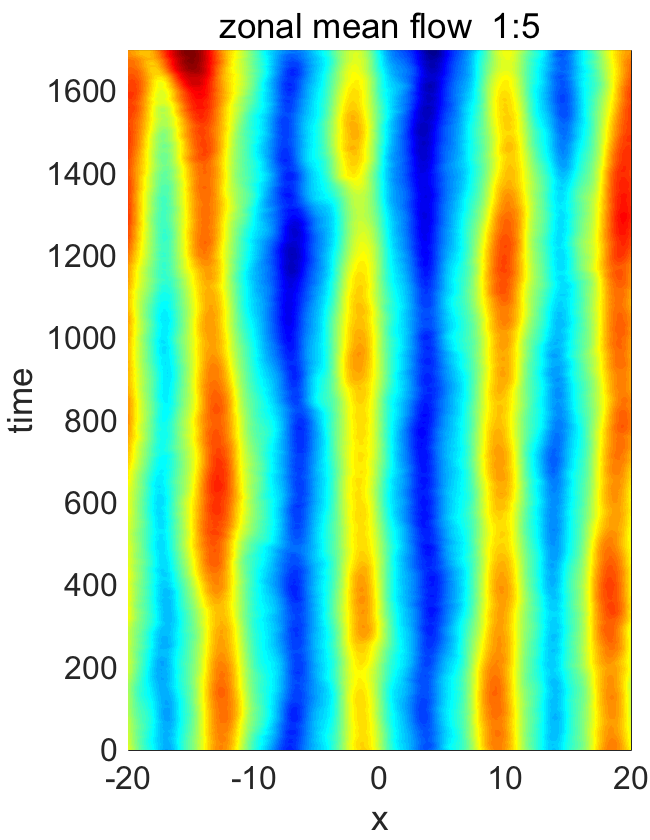}\includegraphics[scale=0.14]{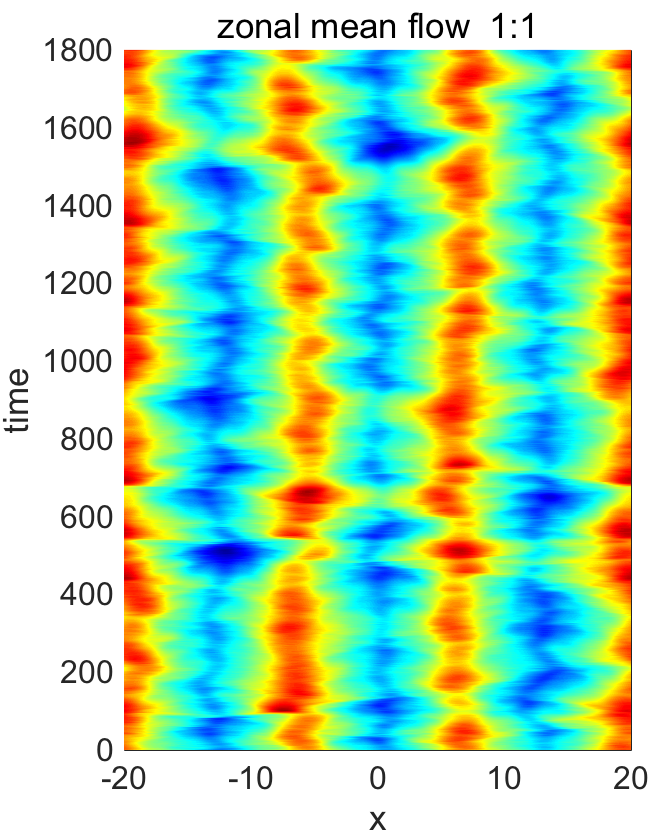}
}

\caption{Time series of the zonal mean flow $v=\partial_{x}\bar{\varphi}$ with
aspect ratios 5:1, 1:5, and 1:1 from simulations of the bHW model.\label{fig:Time-series-v3}}
\end{figure}

We finish this section on the analysis of the flow structures with a comparative study of the time evolution of the jets for different aspect ratios of the computational domain. The time series of the zonal mean flow $v=\partial_{x}\bar{\varphi}$ for the three different aspect ratios
5:1, 1:5, and 1:1 and the two different collisionality regimes $\alpha=0.05$ and $\alpha=0.5$ are shown in Figure \ref{fig:Time-series-v3} for bHW simulations. First, in the regime with strong collisionality, $\alpha=0.05$, the flow field is more homogeneous with strong turbulent structures. Clear zonal jet structure are nevertheless visible, with fast
shifts of the positions of the jets in time. The case with an extended domain in the $x$-direction is particularly interesting, with distinctly new features generated as compared to the original situation with a square domain. The zonal structures indeed combine into large-scale structures with strong bursts the jets evolve in time. These new scales are not captured with the other computational domains.

In the regime with low resistivity, $\alpha=0.5$, a larger number of jets is generated, and the jets are stronger. Once more, the computational domain which is extended in the $x$-direction shows distinctly new dynamics. The multiple zonal jets reorganize into groups of jets, and the groups have bursts which are analogous to the ones seen in the
previous strong turbulence case with $\alpha=0.05$. The emergence of groups at larger scales and the multi-scale nature of the dynamics is unique to this short and wide computational domain, and is due to the interaction of a large number of zonal jets.  

\begin{figure}
\subfloat[mHW $\alpha=0.05,\kappa=0.5$]{\includegraphics[scale=0.14]{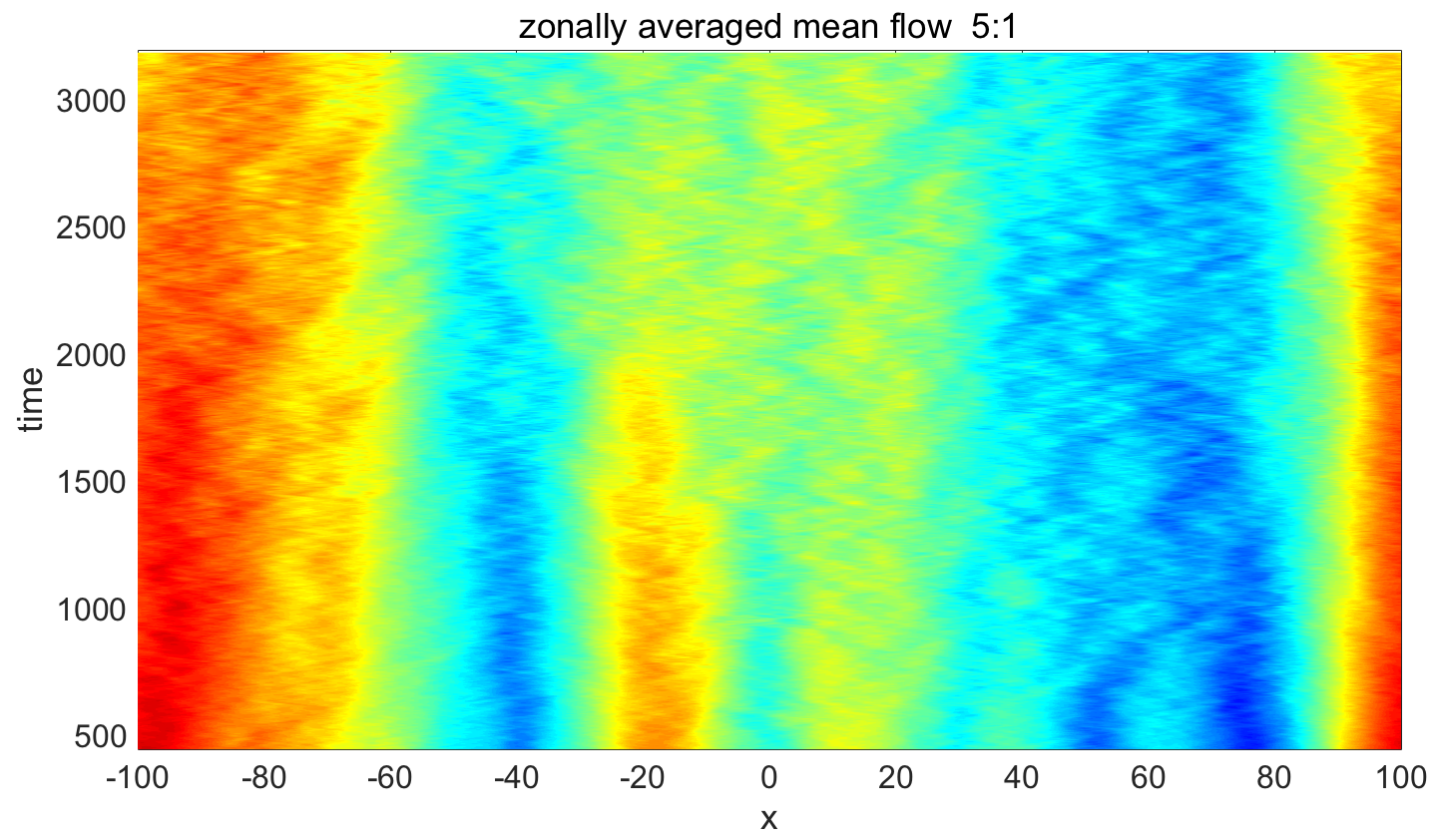}\includegraphics[scale=0.14]{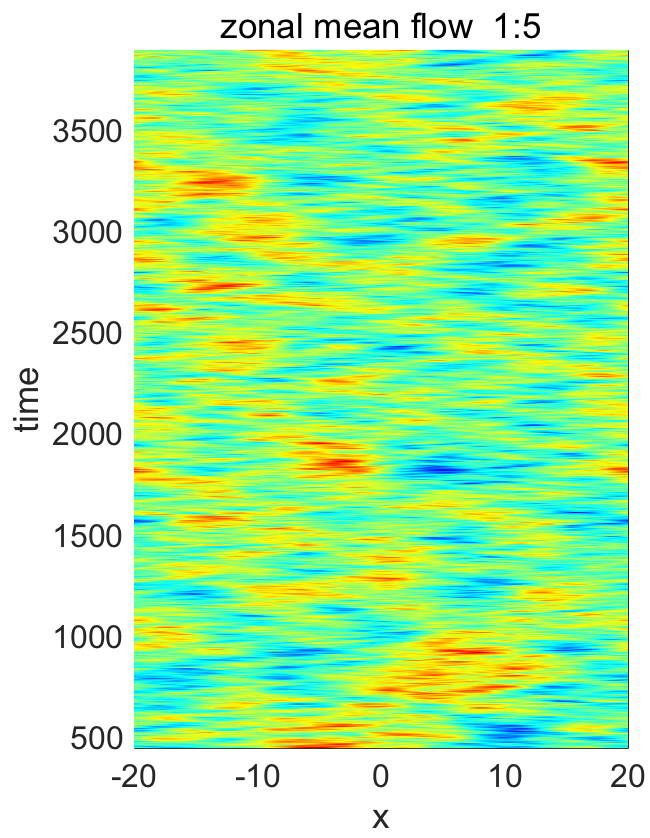}\includegraphics[scale=0.14]{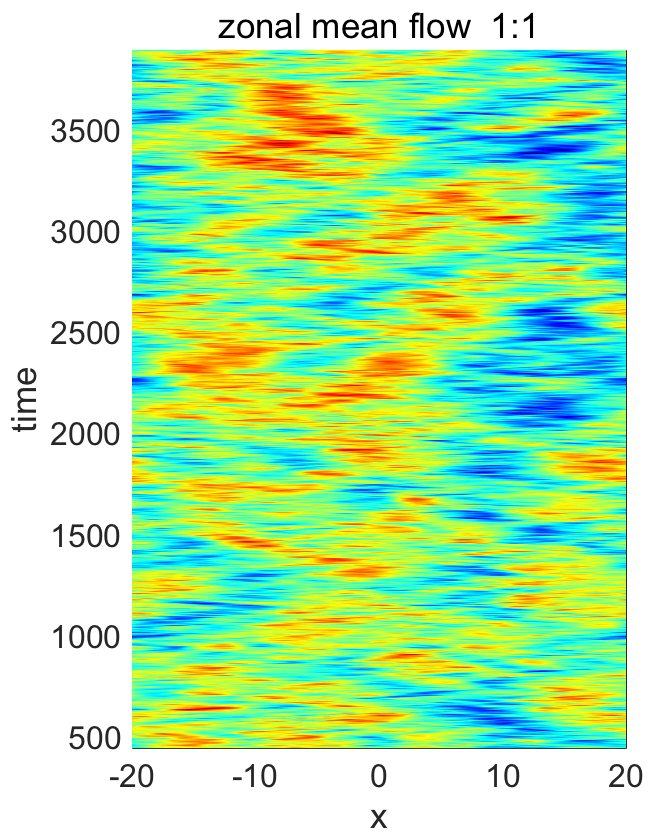}

}

\subfloat[mHW $\alpha=0.5,\kappa=0.5$]{\includegraphics[scale=0.14]{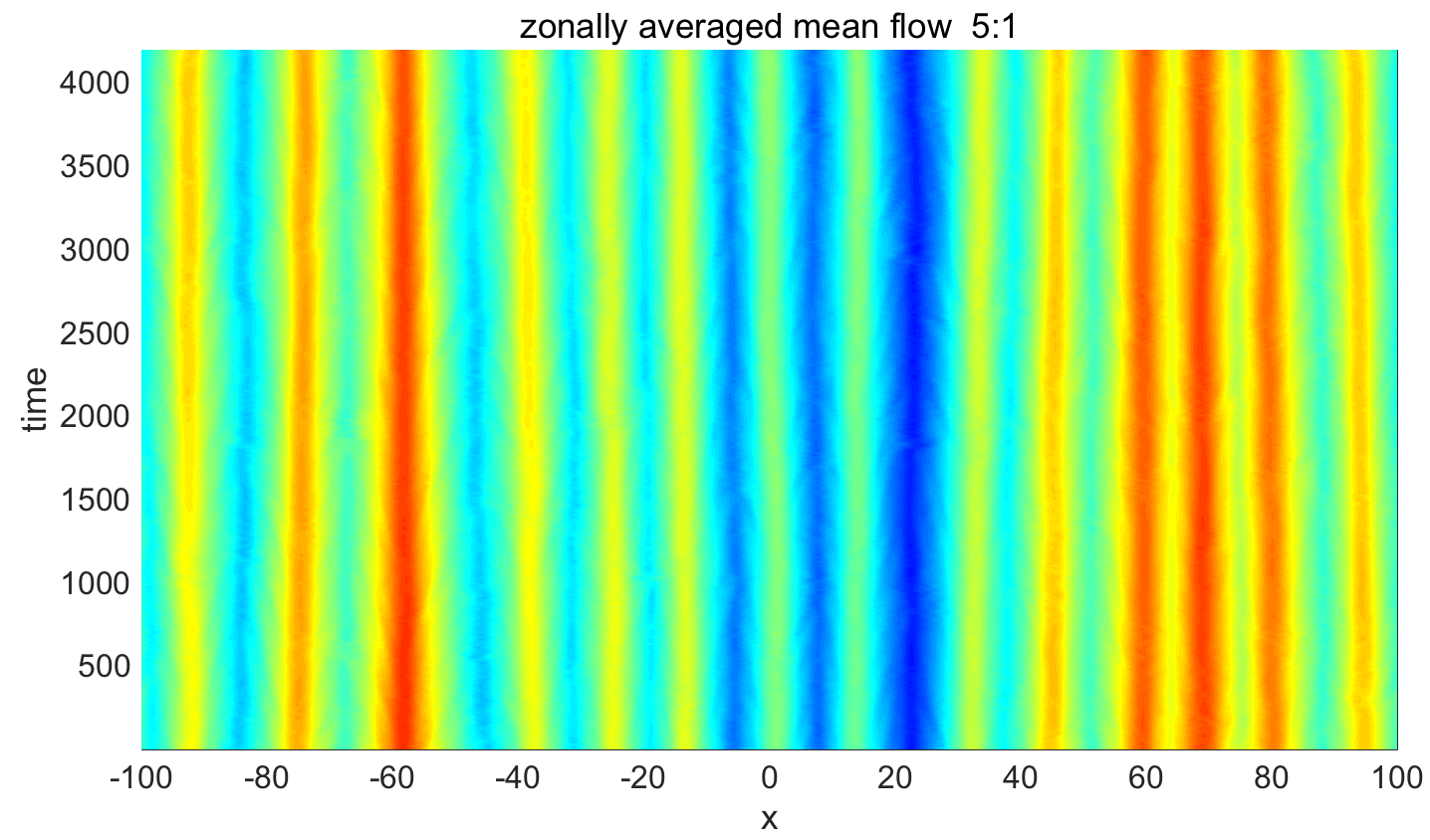}\includegraphics[scale=0.14]{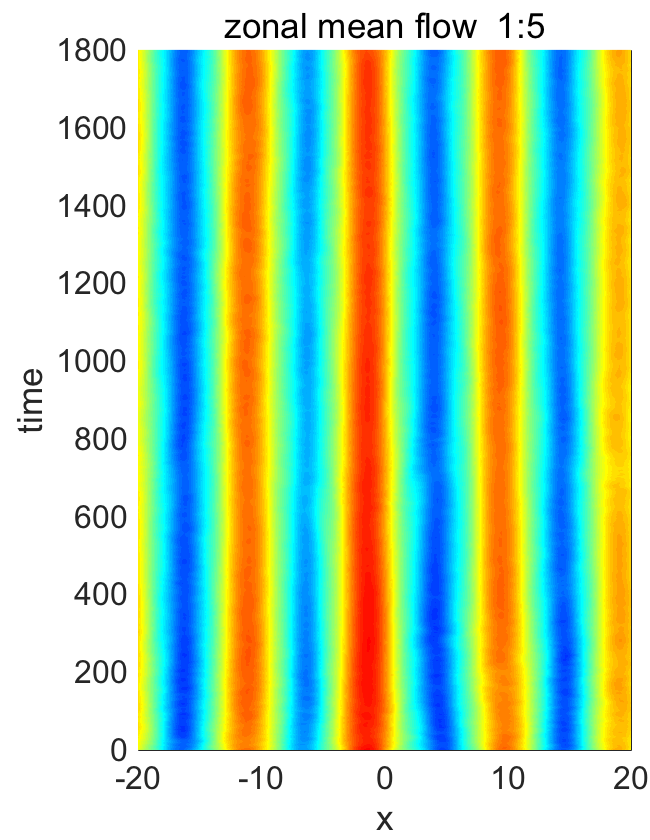}\includegraphics[scale=0.14]{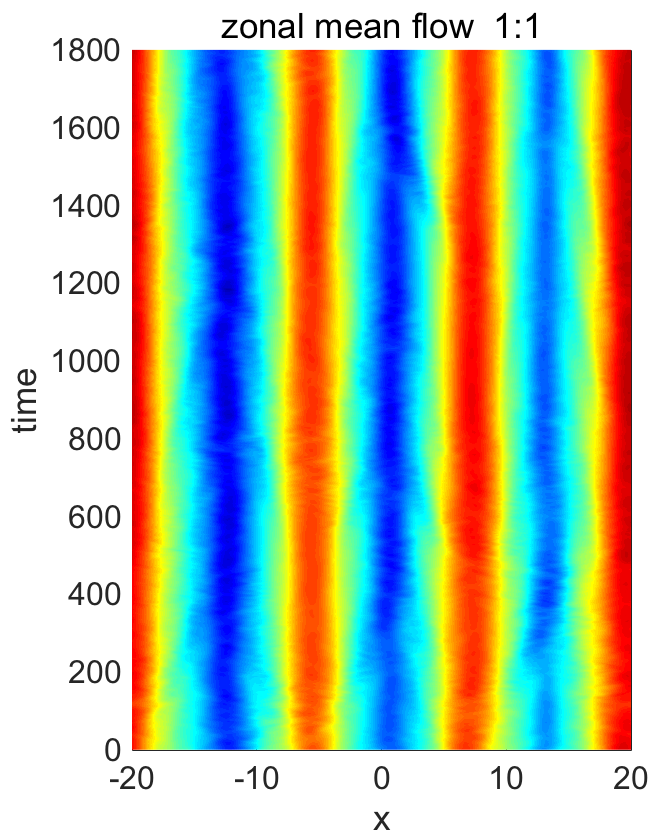}

}

\caption{Time series of the zonal mean flow $v=\partial_{x}\bar{\varphi}$ with
aspect ratios 5:1, 1:5, and 1:1 from simulations of the mHW model.\label{fig:Time-series-v4}}
\end{figure}

It is interesting at this point to compare our results with results from the mHW model in the same configuration, which are shown in Figure \ref{fig:Time-series-v4}.
In the zonal jet dominated regime, for $\alpha=0.5$, there is little variability in the flow pattern, which is consistent with what we observed previously. Interestingly, one cannot discern interactions between distinct jets, even for the case with a short and wide computational domain. In the turbulence dominated regime, for $\alpha=0.05$,
the flow is more homogeneous in space than in the bHW model, except for the domain extended in the $x$-direction, for which distinct, wide zonal structures emerge, as well as interactions between separate jets.

\subsection{Comparison of model statistics with different aspect ratios}

For more quantitative information regarding the modification of the dynamics corresponding to changes in the aspect ratio of the computational domain, we take a statistical viewpoint, as we did previously for the square computational domain. First, we remind the reader that the bHW model has the generalized energy invariant $E=1/2\int_{\Omega}\left(|\nabla\varphi|^2+n^2\right)dV$, where $\Omega$ is the computational domain \citep{majda2018flux}. This motivates us to start the statistic analysis by comparing the statistical energy in the flow kinetic energy $R_{\varphi}=\left\langle \left|\nabla\varphi\right|^{2}\right\rangle _{\mathrm{eq}}$ and in the particle density $R_{n}=\left\langle n^{2}\right\rangle _{\mathrm{eq}}$ for the different aspect ratios and values of the parameter $\alpha$: $\alpha=0.05, 0.2, 0.5$. As before, the symbols $\langle\cdot\rangle_{\mathrm{eq}}$ represent time averages of the quantities of interest once statistical equilibrium has been reached. The results are presented in Table \ref{tab:Comparison-of-total-ene}. One first notices that for all three aspect ratios, as the value of $\alpha$ increases, both $R_{\varphi}$ and $R_{n}$ decrease as more regular flow structures set it, stronger zonal jets are generated. Now, we also observe that the kinetic energy $R_{\varphi}$ is largest for the computational domain extended in the $x$-direction, which is due to the generation of a larger number of jets and the strong interactions between the jets. For the same reason, $R_{\varphi}$ for the square domain case is somewhat larger than $R_{\varphi}$ for the domain elongated in the $y$-direction. These results confirm the visual impression given by Figure \ref{fig:Time-series-v3}. The situation is reversed for the density energy $R_{n}$, which is highest for the domain elongated in the $y$-direction, and smallest for the domain elongated in the $x$-direction.

\begin{table}
{\small{}}%
\begin{tabular}{ccccccccc}
\toprule 
{\small{}bHW} & \multicolumn{2}{c}{{\small{}$\alpha=0.05$}} &  & \multicolumn{2}{c}{{\small{}$\alpha=0.2$}} &  & \multicolumn{2}{c}{{\small{}$\alpha=0.5$}}\tabularnewline
\midrule 
{\small{}$x:y$} & {\small{}$R_{\varphi}$} & {\small{}$R_{n}$} &  & {\small{}$R_{\varphi}$} & {\small{}$R_{n}$} &  & {\small{}$R_{\varphi}$} & {\small{}$R_{n}$}\tabularnewline
\midrule
\midrule 
{\small{}1:1} & {\small{}1.3965} & {\small{}0.9251} &  & {\small{}0.9080} & {\small{}0.1405} &  & {\small{}0.2369} & {\small{}0.0282}\tabularnewline
\midrule 
{\small{}1:5} & {\small{}1.1922} & {\small{}1.7461} &  & {\small{}0.6261} & {\small{}0.5168} &  & {\small{}0.1605} & {\small{}0.0312}\tabularnewline
\midrule 
{\small{}5:1} & {\small{}9.2507} & {\small{}0.8874} &  & {\small{}5.3158} & {\small{}0.1872} &  & {\small{}0.3491} & {\small{}0.0300}\tabularnewline
\bottomrule
\end{tabular}{\small \par}

\caption{Total statistical energy in the flow $R_{\varphi}=\left\langle \int|\nabla\varphi|^{2}\right\rangle $ and in the particle density $R_{n}=\left\langle \int n^{2}\right\rangle $ for different aspect ratios of the computational domain and different
values of the adiabatic resistivity parameter $\alpha$.\label{tab:Comparison-of-total-ene}}
\end{table}


We now turn to the analysis of statistical spectra, as we did in Section \ref{sec:spectra_square}. In order to distinguish the statistics of the zonal and non-zonal modes, we will consider two types of spectra for each quantity. First, we will look at spectra obtained by summing up over all the modes between two adjacent energy shells in integer wavenumbers, $n\leq k\leq n+1$ with $k=\sqrt{k_{x}^{2}+k_{y}^{2}}$, which we call ``radial average", and has a tendency to highlight isotropic dynamics at a given scale. Second, we will construct spectra focused on the zonal modes, i.e. spectra in terms of the wavenumber $k_{x}$, with $k_{y}=0$ fixed. 

\begin{figure}
\subfloat[$\alpha=0.05,\kappa=0.5$]{\includegraphics[scale=0.22]{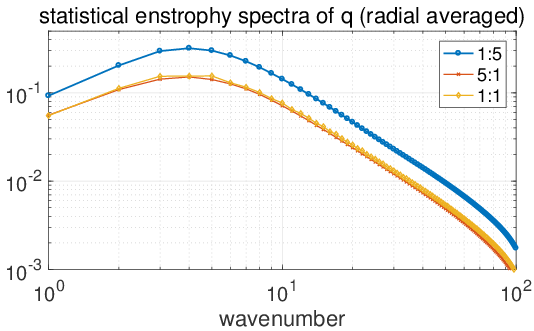}\includegraphics[scale=0.22]{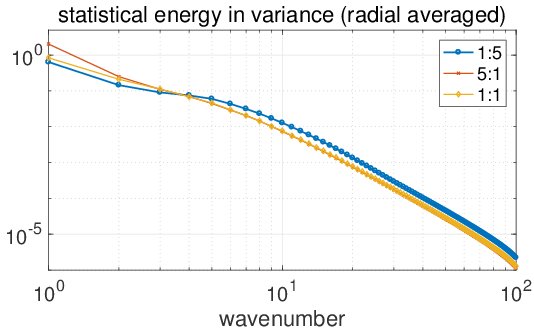}\includegraphics[scale=0.22]{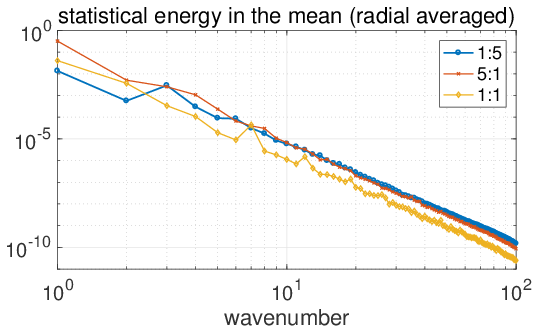}

}

\subfloat[$\alpha=0.5,\kappa=0.5$]{\includegraphics[scale=0.22]{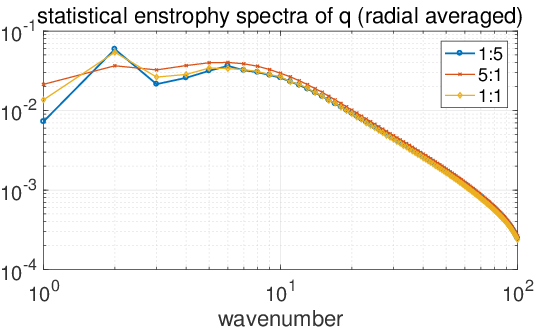}\includegraphics[scale=0.22]{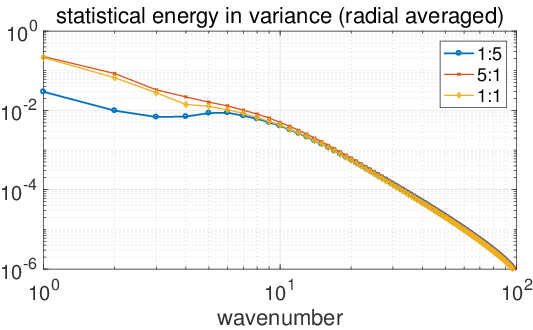}\includegraphics[scale=0.22]{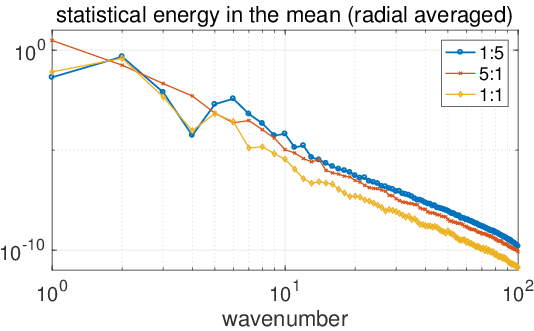}

}

\caption{Radially averaged statistical spectra with aspect ratios 1:5, 5:1,
and 1:1 from the bHW model simulations with parameters $\alpha=0.05,\kappa=0.5$
(top row) and $\alpha=0.5,\kappa=0.5$ (bottom row). From left to right,
we plot the spectra for the statistical enstrophy $\left\langle \left|\hat{q}_{k}\right|^{2}\right\rangle_{\mathrm{eq}} $, the kinetic energy in the variance $k^{2}\left\langle |\hat{\varphi}_{k}^{\prime}|^{2}\right\rangle_{\mathrm{eq}} $, and the kinetic energy in the mean $k^{2}|\left\langle \hat{\varphi}_{k}\right\rangle_{\mathrm{eq}} |^{2}$. \label{fig:Spectra-of-ene_ratio}}
\end{figure}

The radially averaged spectra are shown in Figure \ref{fig:Spectra-of-ene_ratio}, for the total enstrophy total enstrophy $\langle|\hat{q}_{k}|^2\rangle_{\mathrm{eq}}$, with $\hat{q}_{k}=k^{2}\hat{\varphi}_{k}+\hat{n}_{k}$, in the first column, the kinetic energy in the variance $\langle|k\hat{\varphi}_{k}'|^2\rangle_{\mathrm{eq}}$ in the second column, and the kinetic energy in the mean $|\langle k\hat{\varphi}_{k}\rangle_{\mathrm{eq}}|^2$ in the third column, for $\alpha=0.05$ in the top row, and $\alpha=0.5$ in the bottom row, and for the three different aspect ratios for the computational domain considered in this article. We will highlight three key features of this plot. First, in the turbulence dominated regime, $\alpha=0.05$, the statistical enstrophy is much larger in the 1:5 aspect ratio case than in the other two cases. This is due to the much stronger contribution of the density fluctuation to the enstrophy for the computational domain extended in the $y$-direction. Second, the kinetic energy
in the variance in the 1:5 aspect ratio case is always smaller than in the other two cases. This is the signature of more persistent and fixed-in-time jets in the 1:5 aspect ratio case with longer $y$-direction length, in contrast to the other two aspect ratios, which have a stronger jet merging and regeneration dynamics, as we had noted in Figure \ref{fig:Time-series-v3}. Finally, the energy spectra in the mean show that the 5:1 ratio case has larger energy in the mean at large scales than the other two cases. This confirms the self-organization into large multi-jet structures observed in Figure \ref{fig:Time-series-v3}.

\begin{figure}
\subfloat[$\alpha=0.05,\kappa=0.5$]{\includegraphics[scale=0.30]{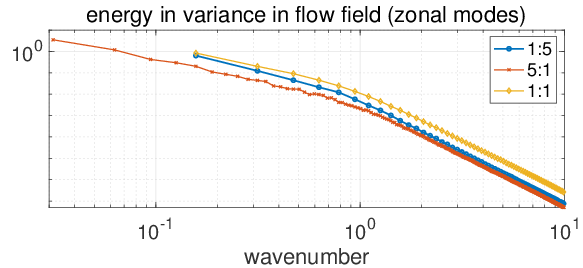}\includegraphics[scale=0.30]{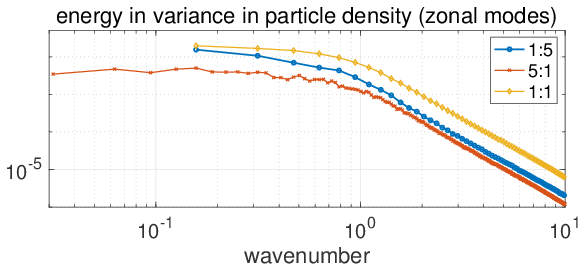}

}

\subfloat[$\alpha=0.5,\kappa=0.5$]{\includegraphics[scale=0.30]{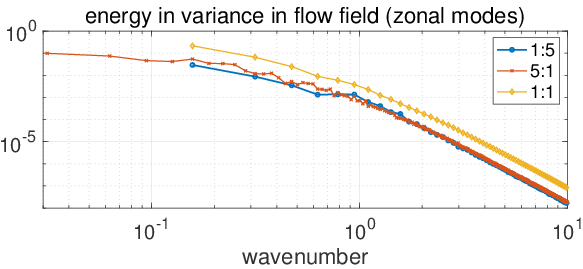}\includegraphics[scale=0.30]{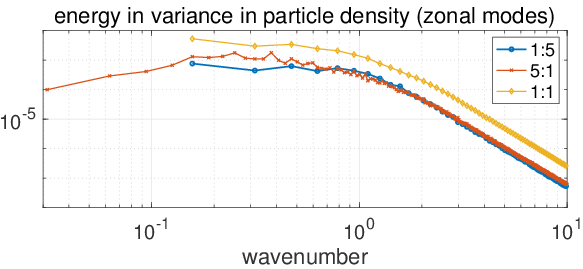}

}

\caption{Zonal statistical spectra with aspect ratios 1:5, 5:1, and 1:1 from
the bHW model simulations with parameters $\alpha=0.05,\kappa=0.5$
(top row) and $\alpha=0.5,\kappa=0.5$ (bottom row). The spectra for the kinetic energy in the variance $k^{2}\left\langle |\hat{\varphi}_{k}^{\prime}|^{2}\right\rangle_{\mathrm{eq}}$ are in the left column,
and the spectra for the energy in the variance of the density $\left\langle|\hat{n}_{k}^{\prime}|^{2}\right\rangle_{\mathrm{eq}} $ are shown in the right column. \label{fig:Spectra-of-ene_ratio-1}}
\end{figure}

We now look at the zonal mode spectra, focusing on the kinetic energy in the variance $k^{2}\langle\left|\hat{\varphi}'_{k}\right|^{2}\rangle_{\mathrm{eq}}$ and the energy in the variance of the density $\langle\left|\hat{n}'_{k}\right|^{2}\rangle_{\mathrm{eq}}$. Our results are shown in Figure \ref{fig:Spectra-of-ene_ratio-1}. Note that because of the nature of the zonal mode spectra, which depend on $k_{x}$ instead of $k$, the spectra for the computational domain extended in the $x$ direction extend to much smaller values of the wavenumber. The main conclusion one can draw regarding the zonal mode spectra is that the energy in the variance is consistently larger for the square computational domain than for the domains extended in either the $x$ or the $y$ direction. This indicates that when the computational domain is extended in either direction, large amounts of energy are distributed to the non-zonal modes to generate stronger fluctuating vortices. We also observe that for the case with a 5:1 aspect ratio, which is elongated in the $x$ direction, the large scales that are unresolved for the other two computational domains have significant amounts of energy, highlighting the large variability at large scales in that case, and interactions between the multiple jets.

\section{Higher-Order Feedbacks in the Statistical Energy Equation\label{sec:Higher-Order-Feedbacks}}

In this final section, we investigate the mechanisms involved in the saturation of the resistive drift instability of the HW models and in the establishment of the statistically stationary state. To do so, we adopt the statistical energy equation framework developed in \citep{majda2018strategies,qi2016predicting} for general turbulent systems, and pay close attention to the third-order moment feedbacks, comparing the mHW and bHW models, as well as different aspect ratios of the computational domain for the bHW model.

\subsection{Third-order moment feedbacks for the statistical energy balance}

\subsubsection{Statistical dynamical equation for the energy matrix }
The statistical energy matrix $R_{k}$ is defined for each spectral mode as
\begin{equation}
R_{k}=\begin{bmatrix}\left\langle k^{2}\left|\hat{\varphi}_{k}\right|^{2}\right\rangle  & \left\langle \hat{\zeta}_{k}\hat{n}_{k}^{*}\right\rangle \\
\left\langle \hat{\zeta}_{k}^{*}\hat{n}_{k}\right\rangle  & \left\langle \left|\hat{n}_{k}\right|^{2}\right\rangle 
\end{bmatrix},\label{eq:stat_ene}
\end{equation}
where $\left\langle \cdot\right\rangle $ stands for the statistical ensemble average of trajectories starting from different initial states. The diagonal entries in $R_{k}$ correspond to the \emph{statistical kinetic energy} $R_{\varphi,k}=\left\langle k^{2}|\hat{\varphi}_{k}|^{2}\right\rangle =-\left\langle \hat{\zeta}_{k}\hat{\varphi}_{k}\right\rangle $
and \emph{statistical energy in density} $R_{n,k}=\left\langle |\hat{n}_{k}|^{2}\right\rangle $. The off-diagonal element, $\left\langle \hat{\zeta}_{k}^{*}\hat{n}_{k}\right\rangle =-\left\langle k^{2}\hat{\varphi}_{k}^{*}\hat{n}_{k}\right\rangle $,
represents the interactions between the flow vorticity and density.

The statistical equations for each statistical energy matrix $R_{k}$ in the bHW model can be derived from the spectral dynamical equation (\ref{eq:bhw_spectral}) by multiplying the state variables $\left(\hat{\varphi}_{k},\hat{n}_{k}\right)$ on each side of the equations and taking ensemble averages, as explained in \citep{majda2018strategies} for similar systems of equations. Doing so, the statistical equations for the $R_{k}$ can be written in the following abstract form, which separates the linear operators $L_{k},D_{k}$ and nonlinear feedback $Q_{F}$:
\begin{equation}
\frac{dR_{k}}{dt}=\left(L_{k}+D_{k}\right)R_{k}+R_{k}^{*}\left(L_{k}^{*}+D_{k}^{*}\right)+Q_{F,k}.\label{eq:eqn_stat_R}
\end{equation}
The linear effects are contained in the two operators $L_{k}$ and
$D_{k}$, where $L_{k}$ depends on the parameters $\alpha,\kappa$,
and $D_{k}$ is a negative-definite operator corresponding to the dissipation effects:
\[
L_{k}=-\left(1-\delta_{k_{y},0}\right)\alpha k^{-2}\begin{bmatrix}1 & 1\\
1 & k^{2}
\end{bmatrix}+i\begin{bmatrix}0 & 0\\
\kappa k_{y} & 0
\end{bmatrix},\quad D_{k}=-D\begin{bmatrix}k^{2}\\
 & k^{2}
\end{bmatrix}.
\]
The first term in $L_{k}$ comes from the resistive coupling term in the HW model, and only applies to the fluctuation modes $k_{y}\neq0$ in both the mHW and the bHW models. The second term in $L_{k}$ is a skew-symmetric matrix coming from the background density gradient. Now, all the higher order feedbacks are found in the nonlinear flux term $Q_{F}$ which includes third-order
moments due to the nonlinear terms in the system, that is,
\[
\begin{aligned}Q_{F,k}= & \begin{bmatrix}\left\langle \hat{\varphi}_{k}^{*}J\left(\varphi,\Delta\varphi-\tilde{n}\right)_{k}\right\rangle +\left\langle \hat{\varphi}_{k}J\left(\varphi,\Delta\varphi-\tilde{n}\right)_{k}^{*}\right\rangle  & \left\langle \hat{n}_{k}^{*}J\left(\varphi,\Delta\varphi-\tilde{n}\right)_{k}\right\rangle +\left\langle k^{2}\hat{\varphi}_{k}J\left(\varphi,n\right)_{k}^{*}\right\rangle \\
\left\langle \hat{n}_{k}J\left(\varphi,\Delta\varphi-\tilde{n}\right)_{k}^{*}\right\rangle +\left\langle k^{2}\hat{\varphi}_{k}^{*}J\left(\varphi,n\right)_{k}\right\rangle  & \left\langle \hat{n}_{k}^{*}J\left(\varphi,n\right)_{k}\right\rangle +\left\langle \hat{n}_{k}J\left(\varphi,n\right)_{k}^{*}\right\rangle 
\end{bmatrix}\\
 & \qquad+\left(1-\delta_{k_{y},0}\right)\begin{bmatrix}\left\langle \hat{\varphi}_{k}^{*}J\left(\varphi,n\right)_{k}\right\rangle +\left\langle \hat{\varphi}_{k}J\left(\varphi,n\right)_{k}^{*}\right\rangle  & \left\langle \hat{n}_{k}^{*}J\left(\varphi,n\right)_{k}\right\rangle \\
\left\langle \hat{n}_{k}J\left(\varphi,n\right)_{k}^{*}\right\rangle  & 0
\end{bmatrix}.
\end{aligned}
\]
We highlight the fact that the second term in the expression above is only nonzero for non-zonal modes. The difference between zonal and non-zonal modes is due to the removal of the zonal mean density $\overline{n}$ in the definition of the potential vorticity in the bHW model. We also note that for each wavenumber $k$, the quadratic terms $J\left(\varphi,f\right)_k$ introduce contributions from other scales through the triad interactions between modes $k=m+n$.

The statistical dynamics described by equation (\ref{eq:eqn_stat_R}) can be interpreted as follows. The linear operator $L_{k}$ has positive eigenvalues corresponding to linearly unstable directions, as analyzed in Section \ref{sub:Stable-and-unstable}. The energy-conserving nonlinear flux $Q_{F}$ plays the crucial role of mediating the growth of energy in the linearly unstable subspace through the negative eigenvalues of $Q_{F}$, and of transferring the excess energy to the stable subspace, where energy is strongly dissipated, through the positive eigenvalues of $Q_{F}$ \citep{majda2018strategies}. This process ensures that the total statistical energy in the system is balanced, and reaches a statistical equilibrium.

\subsubsection{Calculating the higher-order feedbacks from equilibrium statistics}

In general, solving the statistical equation (\ref{eq:eqn_stat_R}) is very computationally intensive, because the direct numerical evaluation of the nonlinear flux term $Q_{F}$ at each time step requires a Monte-Carlo approach with a large ensemble size. This method is too computationally expensive even for a moderate number of wavenumbers kept in the simulations, due to the inclusion of many triad interactions through the Jacobians.

One may however still gain major insights in the mechanisms of interest by considering that statistical equilibrium has been reached, so that the statistical energy $R_{k}^{\mathrm{eq}}$ is invariant in time. In that case, the third-order moment feedbacks are balanced by the equilibrium lower moments using the steady-state version of Eq.(\ref{eq:eqn_stat_R}), in which the time derivative on the left hand side vanishes:
\[
Q_{F}^{\mathrm{eq}}=-\left(L_{k}+D_{k}\right)R_{k}^{\mathrm{eq}}-R_{k}^{\mathrm{eq}*}\left(L_{k}^{*}+D_{k}^{*}\right).
\]
Observe that for each wavenumber $\left(k_{x},k_{y}\right)$, the $2\times2$ nonlinear flux matrix $Q_{F,k}$ has two eigenvalues corresponding to the two eigendirections in $\left(\hat{\varphi}_{k},\hat{n}_{k}\right)$ space. The positive eigenvalues of $Q_{F,k}$ represent the inflow of energy at this scale through the nonlinear energy transfer; in contrast, the negative eigenvalues represent the stabilizing nonlinear effects of removing the excess energy due to the linear instability.

\begin{figure}
\subfloat{\includegraphics[scale=0.24]{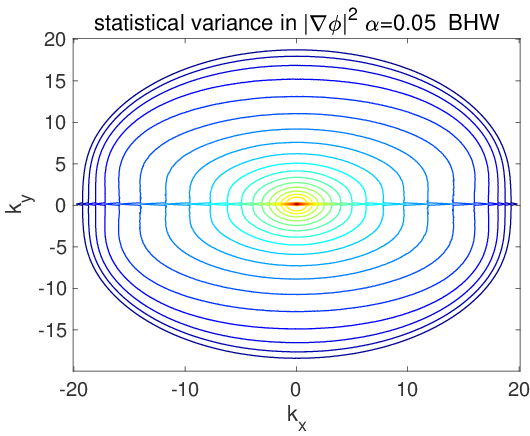}\includegraphics[scale=0.24]{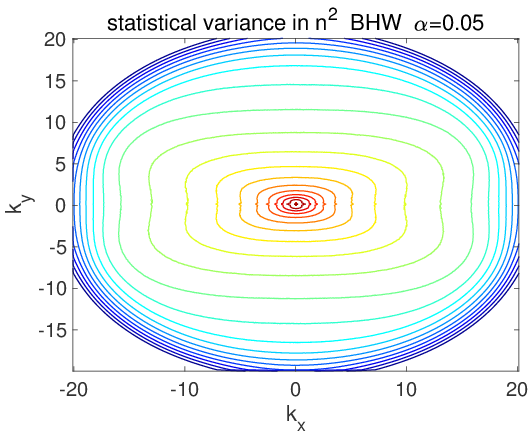}\includegraphics[scale=0.24]{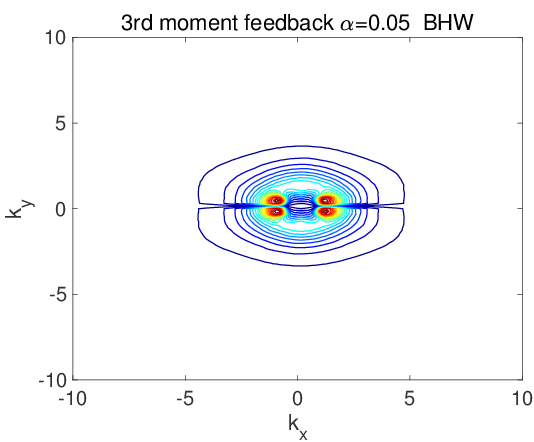}}

\subfloat{\includegraphics[scale=0.24]{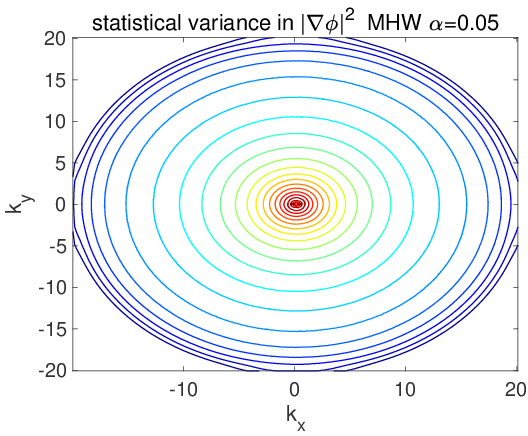}\includegraphics[scale=0.24]{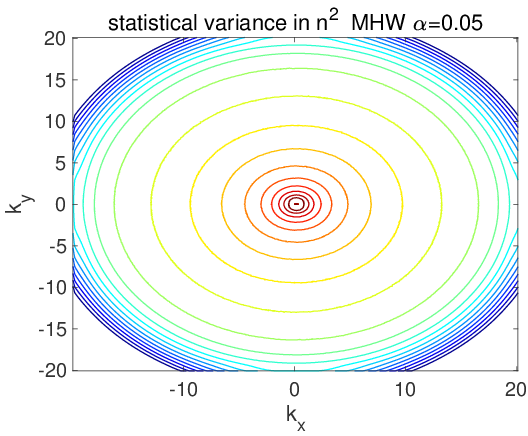}\includegraphics[scale=0.24]{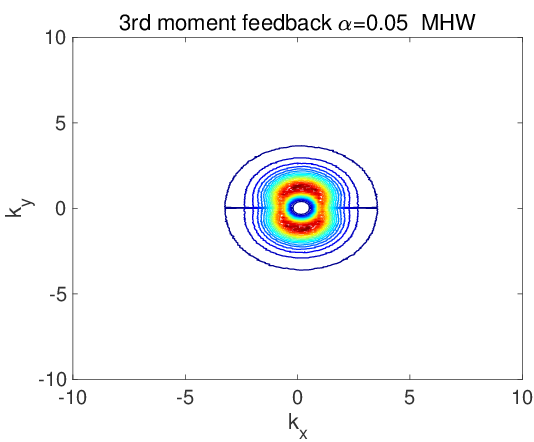}}

\caption{Logarithm of the equilibrium second-order moments $R_{\varphi}^{\mathrm{eq}}$ (Left column) and $R_{n}^{\mathrm{eq}}$ (Middle column), first eigenvalue of $Q_{F}^{\mathrm{eq}}$ (Right column). The plots in the top row were obtained for the bHW model and plots in the bottom row are for the mHW model. All plots were obtained for $\alpha=0.05$.\label{fig:Second-moment1}}
\end{figure}

The first two columns of Figure \ref{fig:Second-moment1} show the logarithm of the statistical variances $R_{\varphi}^{\mathrm{eq}}$ and $R_{n}^{\mathrm{eq}}$ in the spectral domain for the bHW model in the top row, and the mHW model in the bottom row, for the turbulence dominated regime with $\alpha=0.05$. We take the logarithm to emphasize the values at smaller scales and thus visualize contours throughout the plotting window. These figures highlight a central difference between the bHW model and the mHW model. The contours for the bHW model are highly anisotropic, with much larger variances along the zonal modes, with $k_{y}=0$. In contrast, the contours for the mHW model are isotropic.

In the third column of Figure \ref{fig:Second-moment1}, we plot the contours of the first eigenvalue of $Q_{F}^{\mathrm{eq}}$ in the spectral domain, which we compute from Eq.(\ref{eq:eqn_stat_R}), and which is always positive. As explained previously, a positive eigenvalue corresponds to the injection of energy at the given scale from the third-order interactions between different scales. Comparing these results with the linear instability plots shown in Figure \ref{fig:Linear-instability}, we notice that the largest third order moment feedbacks in the bHW model take place near the zonal modes with $k_{y}=0$, and away from the fastest growing linearly unstable modes. This illustrates the nonlinear energy mechanism which transfers the increased energy away from the linearly unstable drift modes to the linearly stable subspace, and leads to the generation of robust zonal jet structures. In contrast, the third-order moments in the mHW model are much more isotropic in this turbulence dominated regime with $\alpha=0.05$, with large values in regions in which linearly unstable drift modes have the fastest growth. This explains the fully homogeneous turbulent flow field obtained in the mHW model and shown in Figure \ref{fig:Snap_vort}. We observe that for both the bHW model and the mHW model, the third order moment feedbacks have a non-trivial structure, which is a signature of non-Gaussian statistical steady-states, as we had already noted in Section \ref{sub:Probability-density-functions}.

\begin{figure}
\includegraphics[scale=0.27]{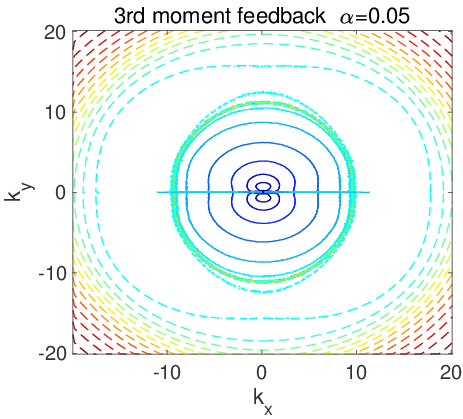}\includegraphics[scale=0.27]{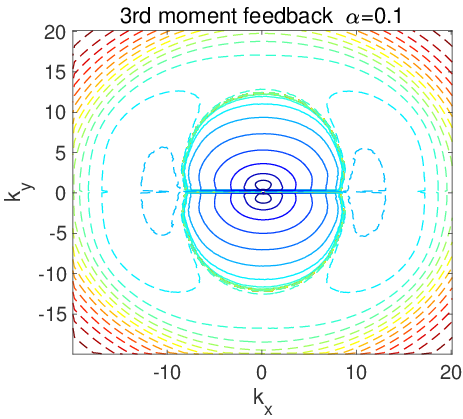}\includegraphics[scale=0.27]{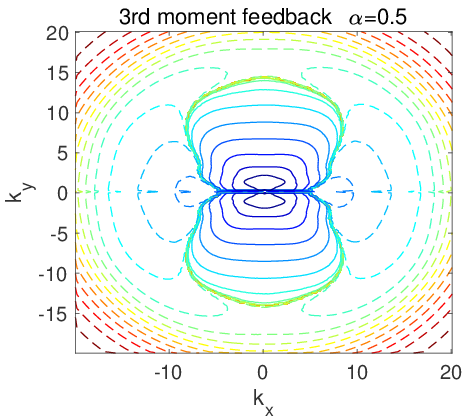}

\caption{Logarithm of the magnitude of the second eigenvalue of $Q_{F}^{\mathrm{eq}}$ in the spectral domain for the bHW model simulations with different values of the parameter $\alpha$: $\alpha=0.05$ (left), $\alpha=0.1$ (middle), and $\alpha=0.5$ (right). Solid lines are used for negative values and dashed lines are used for positive values plotted.\label{fig:Third-order-moment-feedbacks-comp}}
\end{figure}

It is enlightening to look at the transition to the zonal flow dominated regime, corresponding to an increase of the parameter $\alpha$, in the light of the third-order moment feedbacks we study in this section. In Figure \ref{fig:Third-order-moment-feedbacks-comp}, we show contours of the logarithm of the absolute value of the second eigenvalue of $Q_{F}^{\mathrm{eq}}$ in the spectral domain, for three different values of $\alpha$: $\alpha=0.05$, $\alpha=0.1$, and $\alpha=0.5$. Unlike the first eigenvalue of $Q_{F}^{\mathrm{eq}}$, the second eigenvalue may be positive or negative. The contours for negative eigenvalues are plotted with solid lines and the contours for positive eigenvalues are plotted with dashed lines. In the plots, we clearly see the transfer of energy to the small scales near the edges of the spectral domain. We also notice that the negative eigenvalues are localized in a region corresponding to large scales, which closely matches the region of strong linear instability: this is the nonlinear stabilization process for the linearly unstable drift waves. Finally, as the value of $\alpha$ increases, we observe that the region with negative eigenvalues is further localized in the strongly drift unstable region, and more zonal modes have positive eigenvalues. This further illustrates the formation of dominant zonal jet structures as $\alpha$ increases. 

\subsection{Higher-order statistical feedbacks with different aspect ratios}

We now want to study the effects of changing the aspect ratio of the computational domain from the point of view of the third-order moment feedbacks we considered in this section. Table \ref{tab:Comparison-3rd-mom} lists the maximum and minimum eigenvalues of $Q_{F}^{\mathrm{eq}}$ for the three different values of $\alpha$ considered above, $\alpha=0.05$, $\alpha=0.1$, and $\alpha=0.5$, and the three different aspect ratios studied in this work, $1:1$, $5:1$, $1:5$. We find that the computational domains elongated in the $x$ or $y$ directions consistently have stronger third-order interactions, both positive and negative, than the square computational domain
case. Furthermore, the third-order moment feedbacks become stronger as the plasma resistivity increases. This is consistent with the more turbulent dynamics observed for small $\alpha$. In order to provide an energy scale, we also show the maximum value of the variance $R_{\max}$. The range of the third-order moment feedbacks is comparable in magnitude to the magnitude of the variance. This implies that the nonlinear flux term $Q_{F}$ plays a crucial role in
establishing the final dynamics of the statistical equation (\ref{eq:eqn_stat_R}). This is the reason why proper estimation for the higher order statistics is critical for successfully modeling turbulent
systems \citep{majda2016introduction,majda2018strategies,qi2016low}.

\begin{table}
{\scriptsize{}}%
\begin{tabular}{cccccccccccc}
\toprule 
{\scriptsize{}BHW} & \multicolumn{3}{c}{{\scriptsize{}$\alpha=0.05$}} &  & \multicolumn{3}{c}{{\scriptsize{}$\alpha=0.1$}} &  & \multicolumn{3}{c}{{\scriptsize{}$\alpha=0.5$}}\tabularnewline
\midrule 
{\scriptsize{}$x:y$} & {\scriptsize{}$Q_{F,\max}$} & {\scriptsize{}$Q_{F,\min}$} & {\scriptsize{}$R_{\max}$} &  & {\scriptsize{}$Q_{F,\max}$} & {\scriptsize{}$Q_{F,\min}$} & {\scriptsize{}$R_{\max}$} &  & {\scriptsize{}$Q_{F,\max}$} & {\scriptsize{}$Q_{F,\min}$} & {\scriptsize{}$R_{\max}$}\tabularnewline
\midrule
\midrule 
{\scriptsize{}1:1} & {\scriptsize{}$0.32\!\!\times\!\!10^{-3}$} & {\scriptsize{}-0.0058} & {\scriptsize{}0.021} &  & {\scriptsize{}$0.79\!\!\times\!\!10^{-4}$} & {\scriptsize{}-$0.64\!\!\times\!\!10^{-3}$} & {\scriptsize{}0.0064} &  & {\scriptsize{}$0.11\!\!\times\!\!10^{-4}$} & {\scriptsize{}-$0.30\!\!\times\!\!10^{-4}$} & {\scriptsize{}0.0030}\tabularnewline
\midrule 
{\scriptsize{}1:5} & {\scriptsize{}$2.7\!\!\times\!\!10^{-3}$} & {\scriptsize{}-0.0678} & {\scriptsize{}0.016} &  & {\scriptsize{}$4.4\!\!\times\!\!10^{-4}$} & {\scriptsize{}-$3.6\!\!\times\!\!10^{-3}$} & {\scriptsize{}$1.2\!\!\times\!\!10^{-3}$} &  & {\scriptsize{}$4.6\!\!\times\!\!10^{-4}$} & {\scriptsize{}-$4.1\!\!\times\!\!10^{-3}$} & {\scriptsize{}$5.1\!\!\times\!\!10^{-4}$}\tabularnewline
\midrule 
{\scriptsize{}5:1} & {\scriptsize{}$1.6\!\!\times\!\!10^{-3}$} & {\scriptsize{}-0.0272} & {\scriptsize{}0.005} &  & {\scriptsize{}$5.5\!\!\times\!\!10^{-4}$} & {\scriptsize{}-$4.0\!\!\times\!\!10^{-3}$} & {\scriptsize{}$1.9\!\!\times\!\!10^{-3}$} &  & {\scriptsize{}$1.8\!\!\times\!\!10^{-4}$} & {\scriptsize{}-$1.7\!\!\times\!\!10^{-4}$} & {\scriptsize{}$2.7\!\!\times\!\!10^{-4}$}\tabularnewline
\bottomrule
\end{tabular}{\scriptsize \par}

\caption{Maximum and minimum eigenvalues of the matrix $Q_{F}^{\mathrm{eq}}$ of the third-order moment feedback to the total statistical energy $R$ for different aspect ratios of the computational domain. The domains extended in the $x$ or $y$ direction have enhanced third-order moment interactions between modes. For comparison, we also list the maximum variance $R_{\max}$ among the spectral modes to provide an energy scale.\label{tab:Comparison-3rd-mom}}
\end{table}

\begin{figure}
\includegraphics[scale=0.28]{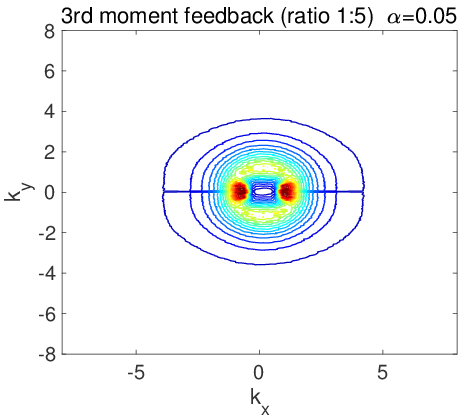}\includegraphics[scale=0.28]{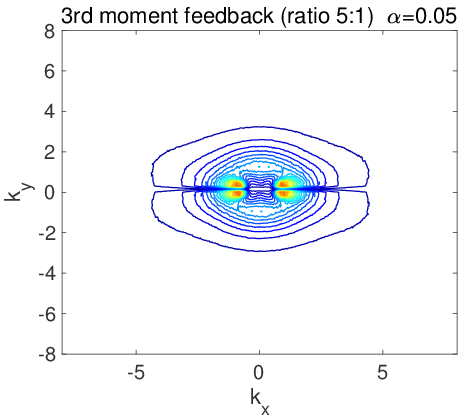}\includegraphics[scale=0.28]{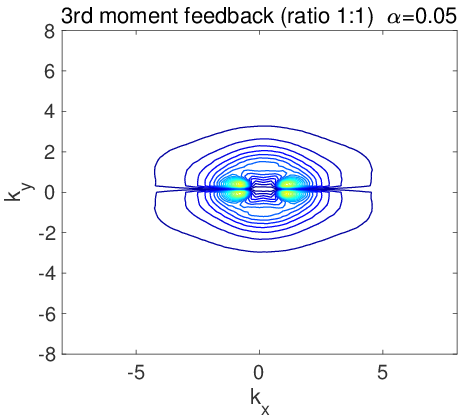}

\caption{Contour of the first eigenvalue of $Q_{F}^{\mathrm{eq}}$ for the aspect ratios 1:5 (left), 5:1 (center), and 1:1 (right) of the computational domain for the bHW model with parameters $\alpha=0.05,\kappa=0.5$. The colormaps are scaled to the same amplitude.\label{fig:Third-order-moment-feedbacks}}
\end{figure}

For a more detailed analysis, in Figure \ref{fig:Third-order-moment-feedbacks} we plot the contours of the first eigenvalue of $Q_{F}^{\mathrm{eq}}$ for the different aspect ratios 1:5, 5:1, and 1:1 and $\alpha=0.05$ in all three cases. To facilitate direct comparison, the colormaps for the three cases are scaled to the same amplitude. The figure confirms that the extended computational domains with aspect ratio 1:5 and 5:1 have stronger higher order interactions than the square computational domain. It also highlights a major difference between the 1:5 and 5:1 cases. The computational domain extended in the $x$ direction has its largest eigenvalues at four symmetric locations with slightly nonzero $k_{y}$, while the third-order moment feedbacks for the computational domain extended in the $y$ direction are strongest for two symmetric regions corresponding to zonal modes with $k_{y}=0$. This is in agreement with the flow dynamics observed in Figure \ref{fig:Time-series-v3}. In the 5:1 case, we had multiple highly fluctuating large scale jets, while in the 1:5 case we had long and dominant zonal jets.

\section{Summary\label{sec:Summary}}
We have studied key characteristics of the drift wave - zonal flow dynamics described by our new flux-balanced Hasegawa-Wakatani model (bHW). This model can be derived from the modified Hasegawa-Wakatani model (mHW) \citep{numata2007bifurcation} by expressing the latter in terms of the potential vorticity and the density, and replacing the standard potential vorticity $\nabla^{2}\varphi-n$ with the \textit{balanced} potential vorticity $q=\nabla^{2}\varphi-\tilde{n}$. This simple modification has significant implications. First, the bHW smoothly converges to the modified Hasegawa-Mima model as one decreases the plasma resistivity, unlike the mHW model. Second, zonal structures are observed for both low and high plasma resistivity. Because of these robust zonal structures, the particle and vorticity fluxes are significantly smaller in the bHW model than in the mHW model in the strongly collisional regime. In order to better understand these differences, we presented a detailed statistical analysis for both the bHW model and the mHW model. We showed that the third-order statistical moments were strongly anisotropic for the bHW model, with a large transfer of energy to the zonal modes. This anisotropy is much less pronounced for the mHW model for high plasma resistivity, explaining the difference in the plasma dynamics.

We have also shown that the dimensions of the computational domain modify the zonal flow dynamics in a major way, particularly so for the bHW model, whose zonal jets have a larger variability than in the mHW model. We found that multi-scale zonal dynamics emerged for computational domains extended in the radial direction, with strong interactions between zonal jets and intermittent multi-jet bursts. When the domain is extended in the poloidal direction instead, the zonal flows have less variability. Our statistical analysis demonstrates that the nature of the third-order moment feedbacks is indeed quite different in the two different cases. For the domain elongated in the poloidal direction, we observe strong energy inputs into the zonal modes, while for the domain elongated in the radial direction the strongest transfers of energy are found to be toward modes which have a small but finite $k_{y}$.

The strongly non-Gaussian features of the bHW turbulence and the major role played by the third-order moment feedbacks suggest that elementary model reduction strategies based on quasi-linear Gaussian approximations are unlikely to provide reliable statistical estimates for this model. The bHW model thus represents an interesting testing ground for more advanced low-order predictive statistical modeling and uncertainty quantification strategies for edge drift turbulence, which could then be applied to more realistic kinetic models of the edge region of magnetic confinement devices.

\section*{Acknowledgements}

This research of the A. J. M. is partially supported by the Office
of Naval Research through MURI N00014-16-1-2161 and DARPA through
W911NF-15-1-0636. D. Q. is supported as a postdoctoral fellow on the
second grant. A.J.C is partially supported by the US Department of Energy, Office of Science, Fusion Energy Sciences under Award Numbers DE-FG02-86ER53223 and DE-SC0012398.

\bibliographystyle{plain}
\bibliography{ref2}

\end{document}